\documentclass[showpacs,aps,pre,preprint,superscriptaddress]{revtex4}

\usepackage{amsmath}
\usepackage{graphicx}
\usepackage{dcolumn}
\usepackage{bm}

\begin{document}

\title{Fluctuating hydrodynamics for dilute granular gases}
\author{J. Javier Brey}
\affiliation{F\'{\i}sica Te\'{o}rica, Universidad de Sevilla,
Apartado de Correos 1065, E-41080, Sevilla, Spain}
\author{P. Maynar}
\affiliation{F\'{\i}sica Te\'{o}rica, Universidad de Sevilla,
Apartado de Correos 1065, E-41080, Sevilla, Spain}
\affiliation{Laboratoire de Physique Th\'eorique (CNRS
  UMR 8627), B\^atiment 210, Universit\'e Paris-Sud, 91405 Orsay Cedex,
  France}
\author{M.I. Garc\'{\i}a de Soria}
\affiliation{F\'{\i}sica Te\'{o}rica, Universidad de Sevilla,
Apartado de Correos 1065, E-41080, Sevilla, Spain}
\affiliation{LPTMS (CNRS UMR 8626), Universit\'e Paris-Sud, Orsay Cedex, F-91405, France}

\date{\today }

\begin{abstract}
Starting from the kinetic equations for the fluctuations and
correlations of a dilute gas of inelastic hard spheres or disks, a
Boltzmann-Langevin equation for the one-particle distribution
function of the homogeneous cooling state is constructed. This
equation is the linear Boltzmann equation with a fluctuating white
noise term. Balance equations for the fluctuating hydrodynamic
fields are derived. New fluctuating forces appear as compared with
the elastic limit. The particular case of the transverse velocity
field is investigated in detail. Its fluctuations can be described
by means of a Langevin equation, but exhibiting two main differences
with the Landau-Lifshitz theory: the noise is not white, and its
second moment is not determined by the shear viscosity. This
shows that the fluctuation-dissipation relations for molecular fluids do not straightforwardly
carry over to inelastic gases. The theoretical predictions are shown to be in good
agreement with molecular dynamics simulation results.
\end{abstract}

\pacs{45.70.Mg, 45.70.Qj, 47.20.Ky} \maketitle

\section{Introduction}
\label{s1} The (modified) nonlinear Boltzmann equation for the one
particle distribution function provides an accurate description of
transport phenomena in a low density gas of inelastic hard spheres or disks
\cite{GyS95,Du00,Go03,PyB03,ByP04}. These particles
are often used to model granular fluids \cite{JNyB96}, specially in
the rapid flow regime \cite{Ca90}. The Boltzmann equation
does not provide any direct information about correlations and
fluctuations in the gas, other that the particle velocity moments. Nevertheless, methods used in the
derivation  of the Boltzmann equation have been extended to obtain
kinetic equations for the equal and different time correlations, in
the same low density approximation. The general idea is that in
order to obtain these equations the needed approximations are the
same as those used to derive the Boltzmann equation itself.

One of the earliest and physically more transparent methods to study
fluctuations is that of Langevin equations. Almost 40 years ago, in
a seminal paper Bixon and Zwanzig \cite{ByZ69} showed how a
Boltzmann-Langevin equation could be constructed by generalizing the
reasonings leading to the Boltzmann equation for molecular gases.
The latter describes the behavior of the average value of the
one-particle distribution function, while the former incorporates
the effects of the fluctuations. As the authors indicated themselves
in the paper, the derivation was based on physical intuition and
analogy. A more systematic derivation of the same result, starting
from first principles, was given in ref. \cite{BDyD89}.

A second approach to the study of correlations in dilute gases makes
use of functional analysis. Its more general result is a kinetic
equation for a generating functional at low density, from which all
multi-point correlations can be obtained by functional
differentiation \cite{MyD83}. A closely related  general scheme for
the study of correlations is the hierarchical method
\cite{EyC81a,EyC81}. The starting point are hierarchies of coupled
equations for the time distribution functions describing the
fluctuations and correlations. Then, the hierarchies are closed by
using the same kind of approximations as needed to derive the
kinetic equation, i.e. the Boltzmann equation in the case of dilute
gases. This method has been recently extended to describe
fluctuations and correlations of dilute inelastic gases in their
simplest state, the homogeneous cooling state (HCS)
\cite{BGMyR04}. As an application, the fluctuations of the total
energy were studied and a good agreement between theory and
simulation results was found \cite{BGMyR04,Vetal06}.

One of the aims of this paper is to translate the above formalism in
terms of kinetic equations for the correlation functions into a
Langevin equation formulation, i.e. to extend the fluctuating
Boltzmann equation to the case of inelastic hard spheres or disks. The
relationship between kinetic equations and the fluctuating Boltzmann
equation has been analyzed in detail in molecular gases
\cite{EyC81,Tr84}. One advantage of the Langevin formulation is that
it is closer to the fluctuating hydrodynamic equations. Actually,
the fluctuating Boltzmann equation for molecular systems has been
shown \cite{ByZ69,Hi70,FyU70} to lead to the same Langevin equations
for the hydrodynamic fields as obtained by Landau and Lifshitz
\cite{LyL66} using thermodynamic fluctuation theory. The noise terms
in these equations are assumed to be white with second moments
determined by the Navier-Stokes transport coefficients of the fluid.
Their expressions are known as fluctuation-dissipation relations of
the second kind \cite{KTyH85}.

The derivation of fluctuating hydrodynamic equations from the
fluctuating Boltzmann equation for inelastic hard particles, will be
also addressed here. Attention will be focussed on a particular
state, the HCS, and on a specific hydrodynamic field, the
transverse component of the velocity. The main conclusion will be
that the fluctuation-dissipation relation for elastic gases can not
be directly extrapolated to inelastic ones, but it needs to be
significantly modified. The second moment of the noise is not determined by
the Navier-Stokes shear viscosity. Moreover, the noise can not be
assumed to be white. These theoretical predictions are in
qualitative and quantitative agreement with molecular dynamics
simulation results.

The consideration of the HCS does not imply by itself that the
results obtained here are not relevant for other states more
accesible experimentally. The HCS plays for
inelastic gases a role similar to the equilibrium state for
molecular gases. In the case of molecular systems, the expressions
of the transport coefficients obtained by linearizing  around
equilibrium are the same as those appearing in the nonlinear
Navier-Stokes equations as predicted by the Chapman-Enskog method
and successfully used in many far from equilibrium problems
\cite{RyL77}. Also, the fluctuation-dissipation relations derived
for near-local-equilibrium states in the original Landau and Lifshitz theory
have proven to be accurate for many other hydrodynamic states
\cite{Ke87}. For dilute gases composed of inelastic hard particles,
the equivalence between the transport coefficients obtained by
linear perturbations of the HCS and by applying the Chapman-Enskog
procedure has also been established \cite{ByD05}. Something similar
might be expected for the fluctuations and correlations.

In the system being considered here, the particles move freely and
independently between consecutive collisions. More specifically, they are not
coupled to any external energy source or thermal bath,
contrary to the driven granular gas models. For
these models, the linear response to an external perturbation
\cite{PByL02} as well as the validity of the Einstein relation
\cite{PByV07} have been investigated by numerical simulations, and
some empirical models have been proposed. It is not evident a direct relation
between the free model considered here and the above driven models.

The plan of the paper is as follows. In Sec. \ref{s2}, the kinetic
equations for the one-time and two-time correlation functions of a
dilute gas in the HCS derived in ref.\ \cite{BGMyR04} are shortly
reviewed. These equations are translated into an equivalent
Boltzmann-Langevin equation for the one particle distribution
function in Sec.\ \ref{s3}. When written in the appropriate
variables, this equation is the linear Boltzmann equation to which a
fluctuating force term is added, similarly to what happens in
molecular elastic gases. An expression for the second moment of the
fluctuating force in terms of the collisional Boltzmann kernel is
derived. In Sec.\ \ref{s4}, the fluctuating hydrodynamic fields are
defined, and balance equations for them are obtained from the
Boltzmann-Langevin equation. They involve formal expressions for the
fluctuating pressure tensor, the fluctuating heat flux, and the
fluctuating cooling rate. In addition, an intrinsically inelastic
fluctuating force shows up in the equation for the energy.

To get a closed description for the hydrodynamic fluctuations,
expressions for the heat flux, the pressure tensor, and the cooling
rate in terms of the fluctuating hydrodynamic fields are needed.
This can be accomplished by means of the Chapman-Enskog procedure. Here
only the case of the transverse
component of the velocity field will be considered. As a
consequence, only the expression for the non-diagonal elements of the
pressure tensor is required. This is computed in Sec. \ref{s5}. The
final result is a Langevin equation, that is the linear macroscopic
equation for the transverse velocity field plus a fluctuating force
term. Therefore, the structure is similar to what one could expect
by extrapolating from the corresponding equation for molecular systems \cite{vNEByO97}.
Nevertheless, the noise term is not white and its second moment is
not given by the usual fluctuation-dissipation relation. It is
verified that the obtained theoretical predictions are in good
agreement with molecular dynamics simulation results. Section
\ref{s7}  contains some general comments and conclusions. Finally,
the appendixes provide some details of the calculations needed to
derive the results presented in the bulk of the paper.

\section{Kinetic equations for the homogeneous cooling state}
\label{s2}

The system considered is a dilute gas of $N$ smooth inelastic hard
spheres ($d=3$) or disks ($d=2$) of mass $m$ and diameter $\sigma$.
The position and velocity of the {\em i}th particle at time $t$ will
be denote by ${\bm R}_{i}(t)$ and ${\bm V}_{i}(t)$, respectively.
The effect of a collision between particles $i$ and $j$ is to
instantaneously modify their velocities according to the rule
\begin{eqnarray}
\label{2.1} {\bm V}_{i} & \rightarrow &{\bm V}_{i}^{\prime} ={\bm
V}_{i} -\frac{1+\alpha}{2} \left( \widehat{\bm \sigma} \cdot {\bm
V}_{ij} \right) \widehat{\bm \sigma}, \nonumber
\\
{\bm V}_{j}  & \rightarrow & {\bm V}_{j}^{\prime} = {\bm V}_{j}
+\frac{1+\alpha}{2} \left( \widehat{\bm \sigma} \cdot {\bm V}_{ij}
\right) \widehat{\bm \sigma},
\end{eqnarray}
where ${\bm V}_{ij}= {\bm V}_{i}-{\bm V}_{j}$ is the relative
velocity, $\widehat{\bm \sigma}$ is the unit vector pointing from
the center of particle $j$ to the center of particle $i$ at contact,
and $\alpha$ is the coefficient of normal restitution. It is defined
in the interval $ 0 < \alpha \leq 1$ and it will considered here as
constant, independent of the velocities of the particles involved in
the collision. A more realistic modeling of granular gases would
require to consider a velocity dependent restitution coefficient
\cite{ByP04}.

Given a trajectory of the system, one-point  and two-point
microscopic densities in phase space at time $t$ are defined by
\begin{equation}
\label{2.2} F_{1}(x_{1},t)=\sum_{j=1}^{N} \delta \left[ x_{1}-
X_{j}(t) \right]
\end{equation}
and
\begin{equation}
\label{2.3} F_{2}(x_{1},x_{2},t)= \sum^{N}_{i} \sum^{N}_{j \neq i}
\delta \left[ x_{1}-X_{i}(t) \right] \delta \left[ x_{2} -X_{j}(t)
\right],
\end{equation}
respectively. Here $X_{i}(t) \equiv \left\{ {\bm R}_{i}(t),{\bm
V}_{i}(t) \right\} $, while the $x_{i} \equiv \left\{ {\bm r}_{i},
{\bm v}_{i} \right\}$ are field variables referring to the
one-particle phase space ($\mu$ space). The density $F_{1}(x_{1},t)$
obeys the equation \cite{EyC81a,BGMyR04}
\begin{equation}
\label{2.4} \left[ \frac{\partial}{\partial t} +  {\bm v}_{1} \cdot \frac{\partial}{\partial {\bm r}_{1}}
\right] F_{1}(x_{1},t)= \int dx_{2} \overline{T}(x_{1},x_{2})
F_{2}(x_{1},x_{2},t),
\end{equation}
with
\begin{equation}
\label{2.6} \overline{T}(x_{i},x_{j}) =  \sigma^{d-1} \int d
\widehat{\bm \sigma}\, \Theta ({\bm v}_{ij} \cdot \widehat{\bm
\sigma}) |{\bm v}_{ij} \cdot \widehat{\bm \sigma}| \left[
\alpha^{-2} \delta ({\bm r}_{ij}-{\bm \sigma}) b_{\bm
\sigma}^{-1}({\bm v}_{i},{\bm v}_{j})-\delta ({\bm r}_{ij}+{\bm
\sigma}) \right],
\end{equation}
where $d \widehat{\bm \sigma}$ is the solid angle element for
$\widehat{\bm \sigma} \equiv {\bm \sigma}/\sigma$, ${\bm r}_{12}
\equiv {\bm r}_{1}- {\bm r}_{2}$, $\Theta$ is the Heaviside step
function, and $b_{\bm \sigma}^{-1}({\bm v}_{1},{\bm v}_{2})$ is an
operator replacing all the functions of ${\bm v}_{1}$ and ${\bm
v}_{2}$ to its right by the same functions of the precollisional
values ${\bm v}^{*}_{1}$ and ${\bm v}^{*}_{2}$ given by
\begin{eqnarray}
\label{2.7} {\bm v}^{*}_{1} \equiv b_{\bm \sigma}^{-1} {\bm v}_{1}=
{\bm v}_{1}-\frac{1+\alpha}{2 \alpha} ( \widehat{\bm \sigma} \cdot
{\bm v}_{12} ) \widehat{\bm \sigma},
\nonumber \\
{\bm v}^{*}_{2} \equiv b_{\bm \sigma}^{-1} {\bm v}_{2}= {\bm
v}_{2}+\frac{1+\alpha}{2 \alpha} ( \widehat{\bm \sigma} \cdot {\bm
v}_{12} ) \widehat{\bm \sigma}.
\end{eqnarray}
It is seen that Eq. (\ref{2.4}) for $F_{1}$ involves the two
particle density $F_{2}$. Actually, it is the first equation of an
infinity hierarchy \cite{BGMyR04}.

The averages of $F_{1}(x_{1},t)$ and $F_{2}(x_{1},x_{2},t)$ over the
initial probability distribution of the system $\rho(\Gamma,0)$,
$\Gamma \equiv \left\{ X_{1}, \ldots, X_{N} \right\}$, are the usual
one-particle and two-particle distribution functions,
\begin{equation}
\label{2.8} f_{1}(x_{1},t) = \langle F_{1}(x_{1},t) \rangle , \quad
f_{2}(x_{1},x_{2},t) = \langle F_{2} (x_{1},x_{2},t) \rangle,
\end{equation}
where the notation
\begin{equation}
\label{2.9} \langle G\rangle \equiv \int d \Gamma\, G(\Gamma)
\rho(\Gamma,0)
\end{equation}
has been employed. Two-time reduced distribution functions can also
defined from the microscopic densities and the initial probability
distribution. The simplest  one is the two-particle two-time
distribution function,
\begin{equation}
\label{2.10} f_{1,1}(x_{1},t ; x_{1}^{\prime},t^{\prime})=  \langle
F_{1}(x_{1},t)F_{1}(x_{1}^{\prime},t^{\prime}) \rangle.
\end{equation}
From the definitions in Eqs. (\ref{2.8}) and (\ref{2.10}) it follows
that
\begin{equation}
\label{2.11} f_{1,1}(x_{1},t;x^{\prime}_{1},t) =\delta
(x_{1}-x^{\prime}_{1})f_{1}(x_{1},t)+f_{2}(x_{1},x^{\prime}_{1},t).
\end{equation}

It is convenient to introduce one-time and two-time correlation
functions by
\begin{equation}
\label{2.12} g_{2}(x_{1},x_{2},t)  \equiv
f_{2}(x_{1},x_{2},t)-f_{1}(x_{1},t) f_{1}(x_{2},t),
\end{equation}
and
\begin{equation}
\label{2.13} h_{1,1}(x_{1},t;x^{\prime}_{1},t^{\prime})  \equiv
f_{1,1}(x_{1},t;x^{\prime}_{1},t^{\prime})
-f_{1}(x_{1},t)f_{1}(x_{1}^{\prime};t^{\prime}),
\end{equation}
respectively. Equation (\ref{2.11}) translates into
\begin{equation}
\label{2.14} h_{1,1}(x_{1},t;x^{\prime}_{1},t)
= \delta (x_{1}-x^{\prime}_{1}) f_{1}(x_{1},t)+  g_{2}(x_{1},x^{\prime}_{1},t).
\end{equation}

In the low density limit, a closed set of kinetic equations for
$f_{1}$, $g_{2}$, and $h_{1,1}$ can be derived \cite{BGMyR04} by
extending the methods developed for molecular gases \cite{EyC81}.
They can be used to analyze the average properties as well as
correlations and fluctuations in arbitrary states of a dilute
granular gas. Here attention will be restricted to a particular
state of a freely evolving granular gas, the so-called homogeneous
cooling state (HCS) \cite{Ha83}. Macroscopically, it is
characterized by a uniform number of particles density $n$, a
vanishing velocity field, and a uniform time-dependent temperature
$T(t)$. It is further defined by the one-particle distribution
function having the scaled form \cite{GyS95}
\begin{equation}
\label{2.15} f({\bm v},t)= n v_{0}^{-d}(t) \chi(c),
\end{equation}
where
\begin{equation}
\label{2.16} v_{0}(t) \equiv \left[\frac{2 T(t)}{m} \right]^{1/2}
\end{equation}
is a thermal velocity and $\chi(c)$ is an isotropic function of the
scaled velocity ${\bm c} \equiv {\bm v}/v_{0}(t)$. The distribution
$\chi (c)$ and the granular temperature $T(t)$ are specified by the
pair of coupled equations
\begin{equation}
\label{2.17}
 \frac{\partial T}{\partial s}=-\zeta_{0}
T(s),
\end{equation}
\begin{equation}
\label{2.18} \frac{\zeta_{0}}{2} \frac{\partial}{\partial  {\bm c}}
\cdot \left( {\bm c} \chi \right) = J_{c}[{\bm c}|\chi].
\end{equation}
In the above expressions,
\begin{equation}
\label{2.19} \zeta_{0}=\frac{(1-\alpha^{2})\pi^{\frac{d-1}{2}}}{2\,
\Gamma \left( \frac{d+3}{2} \right)d} \int d{\bm c}_{1} \int d{\bm
c}_{2}\, c_{12}^{3}\chi({c}_{1}) \chi({c}_{2})
\end{equation}
is the dimensionless cooling rate in the time scale $s$ defined by
\begin{equation}
\label{2.20} s \equiv \int_{0}^{t} dt_{1}
\frac{v_{0}(t_{1})}{\lambda}\, ,
\end{equation}
with $\lambda \equiv (n \sigma^{d-1})^{-1}$, and $J_{c}[{\bm c}|\chi
]$ is the inelastic Boltzmann collision term. Its explicit form is
\begin{equation}
\label{2.21} J_{c}[{\bm c}|\chi]= \int d{\bm c}_{1}\,
\overline{T}_{0}({\bm c},{\bm c}_{1}) \chi({c}) \chi({c}_{1}),
\end{equation}
\begin{equation}
\label{2.22} \overline{T}_{0}({\bm c},{\bm c}_{1})=  \int d
\widehat{\bm \sigma}\, \Theta [({\bm c}-{\bm c}_{1}) \cdot
\widehat{\bm \sigma}] ({\bm c}-{\bm c}_{1}) \cdot \widehat{\bm
\sigma}   \left[ \alpha^{-2} b_{\bm \sigma}^{-1}({\bm c},{\bm
c}_{1}) -1 \right].
\end{equation}

The variable $s$ defined in Eq.\ (\ref{2.20}) is proportional to the accumulated number
of collisions per particle. For thermal velocities, i.e values of $c$ of the
order of unity, a good approximation to the solution of Eqs.\ (\ref{2.17})
and (\ref{2.18}) is provided by the
first Sonine approximation, in which \cite{GyS95,vNyE98}
\begin{equation}
\label{2a.1} \chi(c)= \frac{e^{-c^{2}}}{\pi^{d/2}}\, \left[ 1
+a_{2}(\alpha) S^{(2)} (c^{2}) \right]
\end{equation}
with
\begin{equation}
\label{2a.2}
S^{(2)}(c^{2})= \frac{c^{4}}{4}-\frac{d+2}{2}\, c^{2} +\frac{d(d+2)}{8}
\end{equation}
and
\begin{equation}
\label{2a.3}
a_{2}(\alpha)= \frac{16(1-\alpha)(1-2 \alpha^{2})}{9+24d+(8d-41)\alpha+30 \alpha^{2}
-30 \alpha^{3}}\,.
\end{equation}
In the same approximation
\begin{equation}
\label{2a.4}
\zeta_{0}= \frac{ \sqrt{2} \pi^{(d-1)/2} (1-\alpha^{2})}{\Gamma \left(d/2 \right) d} \left[ 1+ \frac{3 a_{2}(\alpha)}{16} \right].
\end{equation}
A numerically exact solution of Eqs. (\ref{2.27}) and (\ref{2.18}) has been recently reported in
\cite{NBSyG07}. The two-particle one-time correlation function of the HCS is assumed
to have  also a scaled form \cite{BGMyR04}
\begin{equation}
\label{2.23} g_{2}({\bm r}_{12},{\bm v}_{1},{\bm v}_{2},t) = n
\lambda^{-d} v_{0}^{-2d}(t) \widetilde{g} ({\bm l}_{12},{\bm
c}_{1},{\bm c}_{2} ),
\end{equation}
where the scaled length scale ${\bm l} \equiv {\bm r} / \lambda $
has been introduced. The dimensionless correlation $\widetilde{g}$
does not depend on $s$ and obeys the equation
\begin{equation}
\label{2.24}
 \left[ {\bm c}_{12} \cdot \frac{\partial}{\partial
{\bm l}_{12}}  - \Lambda ({\bm c}_{1})-\Lambda ({\bm c}_{2}) \right]
\tilde{g}({\bm l}_{12},{\bm c}_{1},{\bm c}_{2}) = \delta ({\bm
l}_{12}) \overline{T}_{0}({\bm c}_{1},{\bm c}_{2}) \chi( {c}_{1})
\chi({c}_{2}),
\end{equation}
where $\Lambda({\bm c}_{i})$ is the linearized Boltzmann collision
operator \cite{BDyR03},
\begin{equation}
\label{2.25} \Lambda({\bm c}_{i}) \equiv \int d {\bm c}_{3}\,
\overline{T}_{0}({\bm c}_{i},{\bm c}_{3}) (1+P_{i3}) \chi
({c}_{3})-\frac{\zeta_{0}}{2} \frac{\partial}{\partial {\bm c}_{i}}
\cdot {\bm c}_{i}.
\end{equation}
The operator $P_{ij}$ interchanges the labels of particles $i$ and
$j$ of the quantities to its right. For the two-particle two-time
correlation function the scaling reads \cite{BGMyR04}
\begin{equation}
\label{2.26} h_{1,1}(x_{1},t;x_{1}^{\prime},t^{\prime})  =  n
\lambda^{-d} v_{0}^{-d}(t) v_{0}^{-d}(t^{\prime}) \widetilde{h}
({\bm l}_{1}-{\bm l}^{\prime}_{1}, {\bm c}_{1},s- s^{\prime}; {\bm
c}^{\prime}_{1})
\end{equation}
and the kinetic equation is
\begin{equation}
\label{2.27} \left[ \frac{\partial}{\partial s}+{\bm c}_{1} \cdot
\frac{\partial}{\partial {\bm l}_{1}} -\Lambda({\bm c}_{1}) \right]
\tilde{h}({\bm l}_{1}-{\bm l}^{\prime}_{1},{\bm c}_{1},s-s^{\prime};
{\bm c}_{1}^{\prime})=0,
\end{equation}
valid for $s>s^{\prime}>0$. The initial condition for this equation
is
\begin{eqnarray}
\label{2.28} \widetilde{h}({\bm l}_{1}-{\bm l}^{\prime}_{1},{\bm
c}_{1},0; {\bm c}_{1}^{\prime}) & \equiv & \tilde{h}_{1,1}({\bm
l}_{1}-{\bm l}_{1}^{\prime},{\bm c}_{1};{\bm c}_{1}^{\prime}) \nonumber \\
& = & \tilde{g}({\bm l}_{1}-{\bm l}^{\prime}_{1},{\bm c}_{1}, {\bm
c}^{\prime}_{1}) +\delta ({\bm c}_{1}-{\bm c}^{\prime}_{1}) \delta
({\bm l}_{1}-{\bm l}^{\prime}_{1}) \chi({c}_{1}).
\end{eqnarray}
An equation for this distribution follows from Eqs.\ (\ref{2.18})
and (\ref{2.24}),
\begin{equation}
\label{2.29} \left[ {\bm c}_{1} \cdot \frac{\partial}{\partial {\bm
l}_{1}}+{\bm c}_{1}^{\prime} \cdot \frac{\partial}{\partial {\bm
l}_{1}^{\prime}}-\Lambda({\bm c}_{1})-\Lambda({\bm c}_{1}^{\prime})
\right]\widetilde{h}({\bm l}_{1}-{\bm l}^{\prime}_{1},{\bm
c}_{1};{\bm c}^{\prime}_{1}) =\delta ({\bm l}_{1}-{\bm
l}^{\prime}_{1})\widetilde{\Gamma}({\bm c}_{1},{\bm
c}^{\prime}_{1}),
\end{equation}
with
\begin{equation}
\label{2.30} \widetilde{\Gamma} ({\bm c}_{1},{\bm c}^{\prime}_{1}) =
- \left[ \Lambda ({\bm c}_{1})+\Lambda ({\bm c}^{\prime}_{1})
\right] \delta ({\bm c}_{1}-{\bm c}^{\prime}_{1}) \chi ({c}_{1})
+\overline{T}_{0} ({\bm c}_{1},{\bm c}^{\prime}_{1}) \chi ({c}_{1})
\chi({c}^{\prime}_{1}).
\end{equation}

Equations (\ref{2.24}) and  (\ref{2.27}) describe the correlations
between fluctuations in the HCS. They become closed once the
solution to Eqs.\ (\ref{2.17}) and (\ref{2.18}) is known. In the
next section, an alternative and consistent description to that
provided by these kinetic equations will be developed.

\section{Fluctuating Boltzmann equation around the HCS}
\label{s3} Equation (\ref{2.4}) is an exact consequence of the
dynamical equations governing the motion of the particles. The aim
of this section is to approximate it in such a way that give a
closed description of the effective dynamics of a dilute granular
gas in the HCS. To do so, the spatial separation between the centers
of colliding particles will be neglected in the operator
$\overline{T}(x_{1},x_{2})$, and $F_{2}(x_{1},x_{2},t)$ will be approximated by an
effective (Boltzmann) two-particle phase space density at the mesoscopic level
$F_{2}^{B}(x_{1},x_{2},t)$. Moreover, the dimensionless time scale
$s$ and length scale ${\bm l}$ introduced in the previous section
will be used. Then, Eq.\ (\ref{2.4}) becomes
\begin{equation}
\label{3.1} \left( \frac{\partial}{\partial s}+\frac{\zeta_{0}}{2}
\frac{\partial}{\partial {\bm c}_{1}} \cdot {\bm c}_{1} + {\bm
c}_{1} \cdot \frac{\partial}{\partial {\bm l}_{1}} \right)
\widetilde{F}_{1} ({\bm l}_{1},{\bm c}_{1},s) = \int d{\bm c}_{2}\,
\overline{T}_{0} ({\bm c}_{1}, {\bm c}_{2} ) \widetilde{F}_{2}^{B}
({\bm l}_{1},{\bm c}_{1},{\bm l}_{1},{\bm c}_{2},s),
\end{equation}
where dimensionless phase space densities have been defined by
\begin{equation}
\label{3.2} \widetilde{F}_{1} ({\bm l}_{1},{\bm c}_{1},s)=n^{-1}
v_{0}^{d}(t) F_{1}(x_{1},t),
\end{equation}
\begin{equation}
\label{3.3} \widetilde{F}_{2}^{B} ({\bm l}_{1},{\bm c}_{1},{\bm
l}_{2},{\bm c}_{2},s)= n^{-2} v_{0}^{2d}(t) F_{2}^{B}
(x_{1},x_{2},t).
\end{equation}
Comparison of the ensemble average of Eq.\ (\ref{3.1}) with Eq.
(\ref{2.18}) gives the conditions
\begin{equation}
\label{3.4} \langle \widetilde{F}_{1} ({\bm l}_{1},{\bm c}_{1},s)
\rangle_{\text{H}} = \chi({c}_{1}),
\end{equation}
\begin{equation}
\label{3.5} \int d{\bm c}_{2}\ \overline{T}_{0}({\bm c}_{1}, {\bm
c}_{2} ) \langle\widetilde{F}_{2}^{B} ({\bm l}_{1},{\bm c}_{1}, {\bm
l}_{1},{\bm c}_{2},s) \rangle_{\text{H}} = J_{c} \left[ {\bm c} |
\chi\right].
\end{equation}
The subindex $\text{H}$ in the angular brackets indicates that the
ensemble average is taken over the probability distribution for the
HCS.

The deviation of the microscopic density from its average value is
defined by
\begin{equation}
\label{3.6} \delta \widetilde{F}_{1} ({\bm l}_{1},{\bm c}_{1},s)
\equiv \widetilde{F}_{1} ({\bm l}_{1},{\bm c}_{1},s)- \chi
({c}_{1}).
\end{equation}
An evolution equation for this quantity follows by subtracting Eqs.\
(\ref{3.1}) and (\ref{2.18}),
\begin{eqnarray}
\label{3.7}
\left( \frac{\partial}{\partial s} \right. & + & \left.
\frac{\zeta_{0}}{2} \frac{\partial}{\partial {\bm c}_{1}} \cdot {\bm
c}_{1} + {\bm c}_{1} \cdot \frac{\partial}{\partial {\bm l}_{1}}
\right) \delta
\widetilde{F}_{1} ({\bm l}_{1},{\bm c}_{1},s) \nonumber \\
& = &  \int d{\bm c}_{2}\, \overline{T}_{0} ({\bm c}_{1}, {\bm
c}_{2} ) \left[ \widetilde{F}_{2}^{B} ({\bm l}_{1},{\bm c}_{1},{\bm
l}_{1},{\bm c}_{2},s) -\chi({c}_{1})\chi ({c}_{2}) \right].
\end{eqnarray}
The structure of this equations suggests to introduce a cluster
decomposition for $\widetilde{F}_{2}^{B}$ of the form
\begin{equation}
\label{3.8} \widetilde{F}_{2}^{B} ({\bm l}_{1},{\bm c}_{1},{\bm
l}_{1},{\bm c}_{2},s) =  \chi({c}_{1})\chi ({c}_{2}) + \chi({c}_{1})
\delta \widetilde{F}_{1}({\bm l}_{1},{\bm c}_{2},s) + \chi({c}_{2})
\delta \widetilde{F}_{1}({\bm l}_{1},{\bm c}_{1},s)  +
\widetilde{\Phi}_{2}^{B}({\bm l}_{1},{\bm c}_{1},{\bm c}_{2},s).
\end{equation}
This equation defines the microscopic correlation density
$\widetilde{\Phi}_{2}^{B}({\bm l}_{1},{\bm c}_{1},{\bm c}_{2},s)$.
Substitution of its ensemble average in Eq.\ (\ref{3.5}) yields
\begin{equation}
\label{3.9} \int d{\bm c}_{2}\, \overline{T}_{0} ({\bm c}_{1}, {\bm
c}_{2}) \langle \widetilde{\Phi}_{2}^{B}({\bm l}_{1},{\bm
c}_{1},{\bm c}_{2},s) \rangle_{\text{H}}=0.
\end{equation}
Moreover, use of Eq.\ (\ref{3.8}) into Eq.\ (\ref{3.7}) allows to
rewrite the equation in the equivalent form
\begin{equation}
\label{3.10} \left[ \frac{\partial}{\partial s} +{\bm c}_{1} \cdot
\frac{\partial}{\partial {\bm l}_{1}} - \Lambda ({\bm c}_{1}) \right]
\delta \widetilde{F}_{1} ({\bm l}_{1},{\bm c}_{1},s)= \widetilde{S}
( {\bm l}_{1}, {\bm c}_{1},s ),
\end{equation}
where
\begin{equation}
\label{3.11} \widetilde{S} ( {\bm l}_{1}, {\bm c}_{1},s ) \equiv
\int d{\bm c}_{2}\, \overline{T}_{0} ({\bm c}_{1}, {\bm c}_{2})
\widetilde{\Phi}_{2}^{B}({\bm c}_{1},{\bm c}_{2},{\bm l}_{1},s)
\end{equation}
and the operator $\Lambda ({\bm c}_{1})$ was defined in Eq.\ (\ref{2.25}). Equation
(\ref{3.10}) can be interpreted as a fluctuating Boltzmann-Langevin equation
for the one-particle distribution function \cite{ByZ69,Hi70,FyU70},
with the ``noise term'' being $ \widetilde{S} ( {\bm l}_{1}, {\bm c}_{1},s )$. Of course,
this does add any new physical insight by itself in the
understanding  of the starting equation (\ref{3.1}). The relevance and usefulness of this
representation will depend on the properties of the noise term. A
first one follows directly from Eq.\ (\ref{3.9}), that is equivalent
to
\begin{equation}
\label{3.12} \langle \widetilde{S} ({\bm l}_{1}, {\bm c}_{1},s
)\rangle_{\text{H}}=0,
\end{equation}
i.e. the noise has zero average. In the following, other properties
of $\widetilde{S}$ will be derived by requiring consistency with the
results derived in the previous section. Multiplication of Eq.\
(\ref{3.10}) by $\delta \widetilde{F}_{1} ({\bm l}_{1}^{\prime},
{\bm c}^{\prime}_{1},s^{\prime})$ with $s^{\prime}<s$, followed by averaging
gives
\begin{equation}
\label{3.13} \left[ \frac{\partial}{\partial s} +{\bm c}_{1} \cdot
\frac{\partial}{\partial {\bm l}} -  \Lambda ({\bm c}_{1}) \right]
\langle \delta \widetilde{F}_{1} ({\bm l}_{1},{\bm c}_{1},s) \delta
\widetilde{F}_{1} ({\bm l}_{1}^{\prime},{\bm
c}_{1}^{\prime},s^{\prime})\rangle_{\text{H}} = \langle
\widetilde{S} ({\bm l}_{1},{\bm c}_{1},s) \delta \widetilde{F}_{1}
({\bm l}_{1}^{\prime},{\bm
c}_{1}^{\prime},s^{\prime})\rangle_{\text{H}}.
\end{equation}
From the definition of $\delta \widetilde{F}_{1} ({\bm l}_{1},{\bm
c}_{1},s)$ it is easily verified that
\begin{equation}
\label{3.14} \langle \delta \widetilde{F}_{1} ({\bm l}_{1},{\bm
c}_{1},s) \delta  \widetilde{F}_{1} ({\bm l}_{1}^{\prime},{\bm
c}_{1}^{\prime},s^{\prime})\rangle_{\text{H}}  =  n^{-1}
\lambda^{-d} \widetilde{h}({\bm l}_{1}-{\bm l}_{1}^{\prime},{\bm
c}_{1},s-s^{\prime}; {\bm c}_{1}^{\prime}),
\end{equation}
where $\widetilde{h}$ is defined in Eq.\ (\ref{2.26}). Therefore,
consistency of Eqs. (\ref{3.13}) and (\ref{2.27}) implies that
\begin{equation}
\label{3.15} \langle \widetilde{S} ({\bm l}_{1},{\bm c}_{1},s)
\delta \widetilde{F}_{1} ({\bm l}_{1}^{\prime},{\bm
c}_{1}^{\prime},s^{\prime})\rangle_{\text{H}}=0,
\end{equation}
for $s>s^{\prime}$. Since, by hypothesis, the parameters of the
system are such that the HCS is stable, the long time solution of
Eq.\ (\ref{3.10}) is
\begin{equation}
\label{3.16} \delta \widetilde{F}_{1} ({\bm l}_{1}, {\bm c}_{1},s) =
\int_{-\infty}^{s} d\tau\, e^{(s-\tau)L({\bm l}_{1},{\bm c}_{1})}
\widetilde{S} ({\bm l}_{1},{\bm c}_{1}, \tau),
\end{equation}
where the linear operator
\begin{equation}
\label{3.17} L({\bm l}_{1},{\bm c}_{1}) \equiv \Lambda ({\bm
c}_{1})-{\bm c}_{1} \cdot \frac{\partial}{\partial {\bm l}_{1}}
\end{equation}
has been introduced. Using Eq. (\ref{3.16}), it is obtained
\[
\left[ L({\bm l}_{1},{\bm c}_{1})+L ({\bm l}_{1}^{\prime},{\bm
c}_{1}^{\prime}) \right] \langle \delta \widetilde{F}_{1} ({\bm
l}_{1},{\bm c}_{1},s) \delta \widetilde{F}_{1} ({\bm
l}_{1}^{\prime},{\bm c}_{1}^{\prime},s^{\prime})\rangle_{\text{H}}
\]
\[
= -\int_{- \infty}^{s} d\tau\, e^{(s-\tau)L({\bm l}_{1},{\bm
c}_{1})} \langle \widetilde{S} ({\bm l}_{1},{\bm c}_{1},\tau)
\widetilde{S} ({\bm l}_{1}^{\prime},{\bm
c}_{1}^{\prime},s)\rangle_{\text{H}} \]
\begin{equation}
\label{3.18} - \int_{- \infty}^{s} d\tau\, e^{(s-\tau)L({\bm
l}_{1}^{\prime},{\bm c}_{1}^{\prime})} \langle \widetilde{S} ({\bm
l}_{1},{\bm c}_{1},s) \widetilde{S} ({\bm l}_{1}^{\prime},{\bm
c}_{1}^{\prime},\tau)\rangle_{\text{H}}.
\end{equation}
This equation must be compared with Eq.\ (\ref{2.29}), having in
mind Eq. (\ref{3.14}). The time independence of the right hand side
of Eq.\ (\ref{2.29}) prompts to introduce the hypothesis that the
noise term $\widetilde{S}$ is Markovian, and write
\begin{equation}
\label{3.20} \langle \widetilde{S} ({\bm l}_{1},{\bm c}_{1},s)
\widetilde{S} ({\bm l}_{1}^{\prime},{\bm
c}_{1}^{\prime},s^{\prime})\rangle_{\text{H}} = H({\bm c}_{1},{\bm
c}^{\prime}_{1}) \delta ({\bm l}_{1}-{\bm l}_{1}^{\prime}) \delta
(s-s^{\prime}).
\end{equation}
On introduction of this into Eq.\ (\ref{3.18}) and comparison with Eq.\ (\ref{2.29}), it
follows that
\begin{equation}
\label{3.21} H({\bm c}_{1},{\bm c}_{1}^{\prime}) = n^{-1}
\lambda^{-d} \widetilde{\Gamma} ({\bm c}_{1},{\bm c}_{1}^{\prime}),
\end{equation}
with $\widetilde{\Gamma}$ defined in Eq.\ (\ref{2.30}).

The properties  given by Eqs.\ (\ref{3.12}), (\ref{3.15}), and
(\ref{3.20}) guarantee that the description provided by the Langevin
equation (\ref{3.10}) leads to the same expressions for the
two-particle, one-time and two-time correlation functions as the
formulation in terms of reduced distributions functions reviewed in
Sec.\ \ref{s2}.

\section{Fluctuating hydrodynamic fields and balance equations}
\label{s4}

The fluctuating number of particles density, $N ({\bm r},t)$,
momentum density, ${\bm G}({\bm r},t)$, and energy density, $E({\bm
r},t)$, are defined in terms of the microscopic phase space density
as
\begin{equation}
\label{4.1} N({\bm r},t) = \int d{\bm v}\ F_{1}(x,t),
\end{equation}
\begin{equation}
\label{4.2} {\bm G}({\bm r},t) = \int d{\bm v}\, m {\bm v}
F_{1}(x,t),
\end{equation}
\begin{equation}
\label{4.3} E({\bm r},t) = \int d{\bm v}\, \frac{mv^{2}}{2}
F_{1}(x,t).
\end{equation}
Dimensionless deviations from their averages values in the HCS are given by
\begin{equation}
\label{4.4} \delta \rho ({\bm l},s) \equiv \frac{\delta N ({\bm
r},t)}{n} = \int d{\bm c}\, \delta \widetilde{F}_{1} ( {\bm l},{\bm
c},s),
\end{equation}
\begin{equation}
\label{4.5} \delta {\bm \omega} ({\bm l},s) \equiv \frac{\delta {\bm
G}({\bm r},t)}{m n v_{0}(t)} = \int d{\bm c}\, {\bm c} \delta
\widetilde{F}_{1} ( {\bm l},{\bm c},s),
\end{equation}
\begin{equation}
\label{4.6} \delta \epsilon ({\bm l},s) \equiv \frac{2 \delta E
({\bm r},t)}{d n T(t)}\, = \frac{2}{d} \int d{\bm c}\, c^{2} \delta
\widetilde{F}_{1} ( {\bm l},{\bm c},s).
\end{equation}
The quantity $\delta {\bm \omega}({\bm l},s)$ is the dimensionless
velocity field. Balance equations for these fluctuating fields
follow by taking velocity moments in the Langevin-Boltzmann equation
(\ref{3.10}) and using the properties of the noise $\widetilde{S}$.
Some details of the calculations are given in appendix \ref{ap1}.
The resulting equations read
\begin{equation}
\label{4.7}
\frac{\partial}{\partial s}\, \delta \rho ({\bm l},s)+\frac{\partial}{\partial {\bm l}} \cdot \delta
{\bm \omega} ({\bm l},s)=0,
\end{equation}
\begin{equation}
\label{4.8} \left( \frac{\partial}{\partial s} -\frac{\zeta_{0}}{2}
\right) \delta {\bm \omega} ({\bm l},s) + \frac{\partial}{\partial
{\bm l}}\, \cdot \delta {\sf \Pi} ({\bm l},s)=0,
\end{equation}
\begin{equation}
\label{4.9} \left( \frac{\partial}{\partial s}  -  \zeta_{0} \right)
\delta \epsilon ({\bm l},s)  +  \frac{d+2}{d}\,
\frac{\partial}{\partial {\bm l}}\, \cdot \delta {\bm \omega} ({\bm
l},s) + \delta \zeta_{0}({\bm l},s)  +  \frac{2}{d}
\frac{\partial}{\partial {\bm l}}\, \cdot \delta {\bm \phi} ({\bm
l},s) = \widetilde{S}_{\epsilon}({\bm l},s).
\end{equation}
In the above equations, $\delta {\sf \Pi}({\bm l}, s)$ and $\delta
{\bm \phi}({\bm l},s)$ are the fluctuating pressure tensor and heat
flux, respectively. Their definitions in terms of the fluctuating
one-particle distribution function are
\begin{equation}
\label{4.10} \delta {\sf \Pi}({\bm l}, s) = \frac{\delta \epsilon
({\bm l},s)}{2} {\sf I}+ \int d{\bm c}\, {\sf \Delta} ({\bm c})
\delta \widetilde{F}_{1} ( {\bm l},{\bm c},s),
\end{equation}
\begin{equation}
\label{4.11} \delta {\bm \phi}({\bm l},s) = \int d{\bm c}\, {\bm
\Sigma} ({\bm c}) \delta \widetilde{F}_{1} ( {\bm l},{\bm c},s),
\end{equation}
where ${\sf I}$ is the unit tensor of dimension $d$, and
\begin{equation}
\label{4.12} {\sf \Delta}({\bm c}) \equiv {\bm c} {\bm c}-
\frac{c^{2}}{d}\ {\sf I},
\end{equation}
\begin{equation}
\label{4.13} {\bm \Sigma}({\bm c}) \equiv \left( c^{2}-
\frac{d^{2}+2}{2} \right) {\bm c}.
\end{equation}
The term $\delta \zeta_{0} ({\bm l}, s)$ represents the fluctuations
of the cooling rate about its average value in the HCS. Its formal
expressions is
\begin{equation}
\label{4.14} \delta \zeta_{0} ({\bm l},s) =
\frac{(1-\alpha^{2})\pi^{\frac{d-1}{2}}}{ \Gamma \left(
\frac{d+3}{2} \right)d}  \int d{\bm c}_{1} \int d{\bm c}_{2}\,
c_{12}^{3}\chi({c}_{1}) \delta \widetilde{F}_{1} (
{\bm l},{\bm c}_{2},s).
\end{equation}
Finally, $\widetilde{S}_{\epsilon}({\bm l},s)$ is a fluctuating
force term having the properties
\begin{equation}
\label{4.15} \langle \widetilde{S}_{\epsilon}({\bm l},s)
\rangle_{\text{H}} =0
\end{equation}
and
\begin{eqnarray}
\label{4.16} \langle  \widetilde{S}_{\epsilon}({\bm l},s)
\widetilde{S}_{\epsilon}({\bm
l}^{\prime},s^{\prime})\rangle_{\text{H}} & = & n^{-1} \lambda^{-d}
\delta (s-s^{\prime}) \delta ({\bm l}-{\bm l}^{\prime}) \left[ \int
d{\bm c}_{1}\, \int d{\bm c}_{2}\, \chi(c_{1}) \chi(c_{2}) \Phi
({\bm c}_{1},{\bm c}_{2}) \right. \nonumber \\
&& \left. - \frac{8}{d^{2}}\, \zeta_{0} \int d{\bm c}\, c^{4}
\chi(c) \right],
\end{eqnarray}
with $\Phi ({\bm c}_{1},{\bm c}_{2})$ given by Eq.\ (\ref{ap1.16}).
This noise term is intrinsic to the inelasticity of collisions and
has no analogue in normal fluids. Of course, in the elastic limit
$\alpha \rightarrow 1$, $\chi(c)$ becomes a Gaussian and the
fluctuating force $\widetilde{S}_{\epsilon}$ is seen to vanish in
agreement with the well known results for hydrodynamic fluctuations
in molecular fluids \cite{LyL66}. The other main differences between
Eq.\ (\ref{4.9}) and the one for molecular gases is the presence of
the two terms involving the cooling rate, $\zeta_{0}$, and its
fluctuations, $\delta \zeta_{0}$. The presence of these terms is
directly associated with existence of the cooling term in the
macroscopic equation for the average energy
\cite{BDKyS98,ByC01,Go03}.

\section{Langevin equation for the velocity field}
\label{s5} Equations (\ref{4.7})-(\ref{4.9}) do not provide a closed
description of the hydrodynamic fluctuations of a granular gas in
the HCS until $\delta {\sf \Pi}$, $\delta {\bm \phi}$, and $\delta
\zeta_{0}$ are expressed in terms of the fluctuating hydrodynamic
fields. This turns out to be not a simple task, and attention will
be restricted in the following to the equation of the velocity field
$\delta {\bm \omega}({\bm l},s)$, Eq. (\ref{4.8}).

Given two functions $f({\bm c})$ and $g({\bm c})$, their scalar
product is defined as
\begin{equation}
\label{5.1} \langle f|g\rangle \equiv \int d{\bm c}\, \chi^{-1} (c)
f^{+}({\bm c}) g({\bm c}),
\end{equation}
where $f^{+}({\bm c})$ is the complex conjugate of $f({\bm c})$.
Next, a projection operator $\mathcal{P}$ is introduced by
\begin{equation}
\label{5.2} \mathcal{P} f({\bm c}) \equiv \sum_{\beta=1}^{d+2}
\xi_{\beta}({\bm c}) \langle\overline{\xi}_{\beta}|f\rangle.
\end{equation}
Here, the functions $\xi_{\beta}({\bm c}), \beta=1,\ldots,d+2$ are
the eigenfunctions of the linear Boltzmann operator $\Lambda ({\bm
c})$ defined in Eq.\ (\ref{2.25}), corresponding to the hydrodynamic
part of its spectrum. Therefore, they are solutions of the equation
\begin{equation}
\label{5.3} \Lambda ({\bm c}) \xi_{\beta}({\bm c}) = \lambda_{\beta}
\xi_{\beta}({\bm c}).
\end{equation}
Their expressions are \cite{BDyR03,ByD05}
\begin{equation}
\label{5.4} \xi_{1}({\bm c})= \chi(c)+\frac{\partial}{\partial {\bm
c}} \cdot \left[ {\bm c} \chi (c) \right], \quad {\bm \xi}_{2}({\bm
c})=-\frac{\partial \chi(c)}{\partial {\bm c}}, \quad \xi_{3}({\bm
c})= -\frac{\partial}{\partial {\bm c}} \cdot \left[ {\bm c} \chi
(c) \right].
\end{equation}
The associated eigenvalues are found to be
\begin{equation}
\label{5.5} \lambda_{1}=0, \quad \lambda_{2}=\frac{\zeta_{0}}{2}\, ,
\quad \lambda_{3} =-\frac{\zeta_{0}}{2},
\end{equation}
the eigenvalue $\lambda_{2}$ being $d$-fold degenerated. Finally the
functions $\overline{\xi}_{\beta}({\bm c})$ are
\begin{equation}
\label{5.6} \overline{\xi}_{1}({\bm c})=\chi(c), \quad \overline{\bm
\xi}_{2} ({\bm c})={\bm c} \chi(c), \quad \overline{\xi}_{3} ({\bm
c})= \left( \frac{c^{2}}{d} +\frac{1}{2} \right) \chi(c).
\end{equation}
The sets of functions $\{ \xi_{\beta}({\bm c})\}$ and $\{
\overline{\xi}_{\beta}({\bm c})\}$ are seen to have the
biorthogonality property
\begin{equation}
\label{5.7} \langle \overline{\xi}_{\beta} |\xi_{\beta^{\prime}}
\rangle =\delta_{\beta,\beta^{\prime}}\, ,
\end{equation}
$\beta, \beta^{\prime}=1,2,\ldots,d+2$. This guarantees that
$\mathcal{P}$ as defined by Eq.\ (\ref{5.2}) is really a projector
operator, i.e. it verifies $\mathcal{P}^{2} = \mathcal{P}$. It
projects any function of ${\bm c}$ onto the subspace spanned by the
hydrodynamic eigenfunctions of $\Lambda$.

In the following, it will be more convenient to work in the Fourier
representation. The Fourier transform of $\delta \widetilde{F}_{1} ({\bm
l},{\bm c},s)$ is
\begin{equation}
\label{5.8} \delta \widetilde{F}_{1} ({\bm k},{\bm c},s) = \int d{\bm l}\,
e^{-i {\bm k} \cdot {\bm l}} \delta \widetilde{F}_{1} ({\bm l},{\bm
c},s).
\end{equation}
By means of $\mathcal{P}$, $\delta \widetilde{F}_{1} ({\bm k},{\bm c},s)$
can be decomposed into its hydrodynamic
and non-hydrodynamic parts,
\begin{equation}
\label{5.9} \delta \widetilde{F}_{1} ({\bm k},{\bm c},s)= \mathcal{P}
\delta \widetilde{F}_{1} ({\bm k},{\bm c},s) +\mathcal{P}_{\perp}
\delta \widetilde{F}_{1} ({\bm k},{\bm c},s),
\end{equation}
where $\mathcal{P}_{\perp} \equiv 1- \mathcal{P}$. The Fourier
representation of the balance equation for the velocity
fluctuations, Eq. (\ref{4.8}), is
\begin{equation}
\label{5.10} \left( \frac{\partial}{\partial s} -
\frac{\zeta_{0}}{2} \right) \delta {\bm \omega} ({\bm k},s) + i {\bm
k} \cdot \delta {\sf \Pi} ({\bm k},s) =0,
\end{equation}
\begin{equation}
\label{5.11} \delta {\sf \Pi} ({\bm k},s) = \frac{\delta \epsilon
({\bm k},s)}{2}\, {\sf I} +\int d{\bm c}\, {\sf \Delta} ({\bm c})
\delta \widetilde{F}_{1} ({\bm k},{\bm c},s),
\end{equation}
where ${\sf \Delta}({\bm c})$ is defined in Eq.\ (\ref{4.12}).
Getting an explicit expression for $ \delta {\sf \Pi} ({\bm k},s) $
in terms of the fluctuating hydrodynamic fields is the next issue to be
addressed. By direct evaluation, it is easily verified that
\begin{equation}
\label{5.12} \int d{\bm c}\, {\sf \Delta} ({\bm c}) \xi_{\beta}({\bm c})=0,
\end{equation}
$\beta = 1, \ldots, d+2$. Hence Eq. (\ref{5.11}) is equivalent to
\begin{equation}
\label{5.13} \delta {\sf \Pi} ({\bm k},s) = \frac{\delta \epsilon
({\bm k},s)}{2}\, {\sf I} +\int d{\bm c}\, {\sf \Delta} ({\bm c})
\mathcal{P}_{\perp} \delta \widetilde{F}_{1} ({\bm k},{\bm c},s).
\end{equation}
To compute $ \mathcal{P}_{\perp} \widetilde{F}_{1} ({\bm k},{\bm
c},s)$, the operator $\mathcal{P}_{\perp} $ is applied to both sides
of the Fourier transform of the Boltzmann-Langevin equation
(\ref{3.10}),
\begin{equation}
\label{5.14} \left\{ \frac{\partial}{\partial s} -
\mathcal{P}_{\perp} \left[ \Lambda ({\bm c}) -i {\bm k} \cdot {\bm
c} \right] \right\} \mathcal{P}_{\perp} \delta \widetilde{F}_{1}
({\bm k},{\bm c},s) = - \mathcal{P}_{\perp} i {\bm k} \cdot {\bm c}
\mathcal{P} \delta \widetilde{F}_{1} ({\bm k},{\bm c},s) +
\mathcal{P}_{\perp} \widetilde{S}({\bm k},{\bm c},s),
\end{equation}
where use has been made of the property $\mathcal{P}_{\perp} \Lambda
\mathcal{P} =0$, that is a consequence of the fact that
$\mathcal{P}$ projects over a subspace generated by eigenfunctions
of $\Lambda$. The solution of the above equation can be formally
written as
\begin{eqnarray}
\label{5.15} \mathcal{P}_{\perp} \delta \widetilde{F}_{1} ({\bm
k},{\bm c},s) & = & \mathcal{U}({\bm k},{\bm c},s)
\mathcal{P}_{\perp}
\delta \widetilde{F}_{1} ({\bm k},{\bm c},0)+ \int_{0}^{s} ds^{\prime}\, \mathcal{U}({\bm k},{\bm
c},s^{\prime})  \mathcal{P}_{\perp} \left[ - i {\bm k} \cdot {\bm c}
\mathcal{P} \delta \widetilde{F}_{1} ({\bm k},{\bm c},s-s^{\prime})
\right. \nonumber \\
&& \left.+ \widetilde{S}({\bm k},{\bm c},s-s^{\prime}) \right],
\end{eqnarray}
with
\begin{equation}
\label{5.16} \mathcal{U}({\bm k},{\bm c},s) \equiv \exp \left[s
\mathcal{P}_{\perp} L({\bm k},c) \mathcal{P}_{\perp} \right],
\end{equation}
\begin{equation}
\label{5.17} L({\bm k},{\bm c}) \equiv \Lambda ({\bm c}) -i {\bm k}
\cdot {\bm c}.
\end{equation}
Taking again into account that the HCS is assumed to be stable for
the system considered, the first term on the right hand side of
Eq.\, (\ref{5.15}) can be neglected for large enough times $s$.
Moreover, to  derive hydrodynamic equations valid to Navier-Stokes
order, only the first order in $k$ of the pressure tensor is needed.
To this order,
\[
\int_{0}^{s} ds^{\prime}\, \mathcal{U}({\bm k},{\bm c},s^{\prime})
\mathcal{P}_{\perp} \left( - i {\bm k} \cdot {\bm c} \right)
\mathcal{P} \delta \widetilde{F}_{1} ({\bm k},{\bm c},s-s^{\prime})
\]
\[ \rightarrow
\int_{0}^{s} ds^{\prime}\, e^{ s^{\prime} \mathcal{P}_{\perp}
\Lambda ({\bm c})  \mathcal{P}_{\perp}} \mathcal{P}_{\perp} \left( - i {\bm k} \cdot
{\bm c} \right) \mathcal{P} \delta \widetilde{F}_{1} ({\bm k},{\bm
c},s- s^{\prime})
\]
\begin{equation}
\label{5.18} = \int_{0}^{s} ds^{\prime}\,
\mathcal{P}_{\perp} e^{ s^{\prime}\Lambda ({\bm c})} \left( - i {\bm
k} \cdot {\bm c} \right) \mathcal{P} \delta \widetilde{F}_{1} ({\bm
k},{\bm c},s- s^{\prime}).
\end{equation}
Then, for large $s$ Eq.\ (\ref{5.15}) reduces to
\begin{eqnarray}
\label{5.19} \mathcal{P}_{\perp} \delta \widetilde{F}_{1} ({\bm
k},{\bm c},s) & =& \int_{0}^{s} ds^{\prime}\, \mathcal{P}_{\perp}
e^{ s^{\prime}\Lambda ({\bm c})} \left( - i {\bm k} \cdot {\bm c}
\right)  \mathcal{P} \delta
\widetilde{F}_{1} ({\bm k},{\bm c},s- s^{\prime}) \nonumber \\
& & + \int_{0}^{s} d s^{\prime}\,  \mathcal{U}({\bm k},{\bm
c},s^{\prime}) \mathcal{P}_{\perp} \widetilde{S}({\bm k},{\bm
c},s-s^{\prime}), \nonumber \\
\end{eqnarray}
and substitution of this into Eq.\ (\ref{5.13}) yields
\begin{equation}
\label{5.20} \delta {\sf \Pi} ({\bm k},s) = \frac{\delta \epsilon
({\bm k},s)}{2}\, {\sf I}+\delta_{1} {\sf \Pi} ({\bm k},s) + {\sf R}
({\bm k},s),
\end{equation}
where
\begin{equation}
\label{5.21} \delta_{1} {\sf \Pi} ({\bm k},s) =   \int_{0}^{s} d
s^{\prime}\, \int d{\bm c}\, {\sf \Delta}({\bm c}) e^{s^{\prime}
\Lambda ({\bm c})} (-i {\bm k} \cdot {\bm c} )   \mathcal{P} \delta
\widetilde{F}_{1} ({\bm k}, {\bm c},s-
s^{\prime}),
\end{equation}
and
\begin{equation}
\label{5.22} {\sf R}({\bm k},s)= \int_{0}^{s} ds^{\prime} \int d{\bm
c}\, {\sf \Delta} ({\bm c}) \mathcal{U} ({\bm k},{\bm c},
s^{\prime}) \mathcal{P}_{\perp} \widetilde{S} ({\bm k},{\bm
c},s-s^{\prime}).
\end{equation}
Upon writing Eq. (\ref{5.21}), Eq.\ (\ref{5.12}) has been employed
to remove the operator $\mathcal{P}_{\perp}$ appearing in the first
term on the right hand side of Eq. (\ref{5.19}). Because of the
isotropy of the operator $\Lambda ({\bm c})$, only the projection
onto ${\bm \xi}_{2}({\bm c})$  gives a non-vanishing contribution to
the above expression for $\delta_{1} {\sf \Pi}({\bm k},s)$, that can
be simplified to
\begin{eqnarray}
\label{5.23} \delta_{1} {\sf \Pi} ({\bm k},s) & = & \int_{0}^{s} d
s^{\prime}\, \int d{\bm c}\, {\sf \Delta}({\bm c}) e^{s^{\prime}
\Lambda ({\bm c})} (-i {\bm k} \cdot {\bm c} ) {\bm \xi}_{2} ({\bm c}) \cdot \langle\overline{\bm
\xi}_{2}({\bm c}) | \delta \widetilde{F}_{1} ({\bm k},{\bm
c},s-s^{\prime}) \rangle \nonumber \\
& = & \int_{0}^{s} d s^{\prime}\, \int d{\bm c}\, {\sf \Delta}({\bm
c}) e^{s^{\prime} \Lambda ({\bm c})} (-i {\bm k} \cdot {\bm c} )
{\bm \xi}_{2} ({\bm
c}) \cdot \delta {\bm \omega} ({\bm k},s-s^{\prime}) \nonumber \\
& \simeq & \int_{0}^{s} d s^{\prime}\, \int d{\bm c}\, {\sf
\Delta}({\bm c}) e^{s^{\prime} \Lambda ({\bm c})} (-i {\bm k} \cdot
{\bm c} ) e^{-s^{\prime} \zeta_{0}/2} {\bm \xi}_{2} ({\bm c}) \cdot
\delta {\bm \omega} ({\bm k},s).
\end{eqnarray}
In the previous transformations, the definition in Eq. (\ref{4.5})
has been used, and it has been taken into account that to lowest
order in $k$,
\begin{equation}
\label{5.24} \delta \omega ({\bm k},s-s^{\prime})= e^{-s^{\prime}
\zeta_{0}/2}\delta \omega ({\bm k},s),
\end{equation}
according to the balance equation for the fluctuating velocity
field, Eq. (\ref{4.8}). Using again the symmetry of ${\bm
\Lambda}({\bm c})$, it is obtained:
\begin{equation}
\label{5.25} \delta_{1} {\Pi}_{ij} ({\bm k},s) = -i
\widetilde{\eta}(s) \left[ k_{i} \delta \omega_{j} ({\bm k},s) +
k_{j} \delta \omega_{i} ({\bm k},s) - \frac{2}{d}\, \delta_{ij} {\bm k} \cdot \delta
\widetilde{\bm \omega} ({\bm k},s) \right].
\end{equation}
This is the same as the Navier-Stokes expression for the pressure
tensor with the only difference that the average macroscopic
velocity field is substituted by the fluctuating one. It involves
the (time-dependent) dimensionless shear viscosity
$\widetilde{\eta}(s)$ defined by
\begin{equation}
\label{5.26} \widetilde{\eta}(s)  =  \frac{1}{d^{2}+d-2} \sum_{i}
\sum_{j} \int_{0}^{s} d s^{\prime} \int d {\bf c}\, \Delta_{ij}({\bm
c})  e^{s^{\prime} (\Lambda-\frac{\zeta_{0}}{2})}
\xi_{2,i}({\bm c}) c_{j}.
\end{equation}
This expression is equivalent to the one obtained from the nonlinear
Boltzmann equation for inelastic hard spheres or disks by the
Chapman-Enskog method \cite{BDKyS98,ByC01} and also to the
Green-Kubo formulas derived in ref. \cite{DyB02}. Let us remark that
the results obtained here apply in the limit of large $s$. It is in
this limit when hydrodynamics in the usual sense is expected to
apply. If this is true, the shear viscosity in Eq.\ (\ref{5.26})
will become independent of $s$. Although there is no a mathematical
proof of this ``ageing to hydrodynamics'' for granular gases up to
now, numerical evaluation of the right hand side of Eq. (\ref{5.26})
by using the direct Monte Carlo simulation method has shown the
existence of such a limit value \cite{ByR04}. Moreover, the
simulation results for the shear viscosity $\widetilde{\eta}$ are in
good agreement with the expression obtained by evaluating the
Chapman-Enskog result in the first Sonine approximation
\cite{BDKyS98,ByC01},
\begin{equation}
\label{5a.26}\widetilde{\eta}(\alpha)= \left[8
\widetilde{\nu}(\alpha) -\zeta_{0}(\alpha) \right]^{-1},
\end{equation}
\begin{equation}
\label{5b.26} \widetilde{\nu}(\alpha) =  \frac{\pi^{\frac{d-1}{2}}}{2 \sqrt{2} d(d+2) \Gamma \left(
d/2 \right)} (3- 3 \alpha + 2d) (1+\alpha) \left[ 1 - \frac{a_{2} (\alpha)}{32}
\right].
\end{equation}

When Eq.\ (\ref{5.25}) is substituted into Eq.\ (\ref{5.20}) and the
resulting expression is used into  Eq.\
(\ref{5.10}), a Langevin-like equation is obtained for the velocity
field,
\begin{equation}
\label{5.27} \left( \frac{\partial}{\partial s} -
\frac{\zeta_{0}}{2} \right) \delta {\bm \omega} ({\bm k},s)  +
\frac{i}{2}\, \delta \epsilon ({\bm k},s) {\bm k}  +
\widetilde{\eta} \left[ k^{2} \delta {\bm \omega} ({\bm k},s)
+  \frac{d-2}{d}\, {\bm k} \cdot \delta {\bm \omega} ({\bm
k},s) {\bm k} \right] ={\bm W}({\bm k},s),
\end{equation}
with the noise term given by
\begin{equation}
\label{5.28} {\bm W}({\bm k},s)\equiv -i {\bm k} \cdot {\sf R} ({\bm
k},s),
\end{equation}
where the term ${\sf R}({\bm k},s)$ is defined in Eq.\ (\ref{5.22}). It
follows from Eq. (\ref{3.12}) that
\begin{equation}
\label{5.29} \langle{\bm W}({\bm k},s)\rangle_{\text{H}}=0.
\end{equation}
A formal expression for the correlation function of ${\bm W}$ is
obtained directly by using Eq.\ (\ref{3.20}). Its conversion into an
explicit one, valid to Navier-Stokes order and, therefore,
consistent with the left hand side of Eq.\ (\ref{5.27}), will be
carried out in the next section for the particular case of the
transverse component of the velocity field.

\section{The noise term in the equation for the transverse velocity field}
\label{s6} The transverse part of the fluctuating velocity field, $\delta {\bm
\omega}_{\perp} ({\bm k},s)$, is defined by
\begin{equation}
\label{6.1} \delta {\bm \omega}_{\perp} ({\bm k},s) \equiv \delta
{\bm \omega}({\bm k}, s)- \delta {\bm \omega}({\bm k},s) \cdot
\frac{{\bm k}}{k^{2}}\, {\bm k}.
\end{equation}
Its evolution equation can be written down directly from Eq.\
(\ref{5.27}),
\begin{equation}
\label{6.2} \left( \frac{\partial}{\partial s} -\frac{\zeta_{0}}{2}
+ \widetilde{\eta} k^{2} \right) \delta {\bm \omega}_{\perp} ({\bm
k},s) = {\bm W}_{\perp}({\bm k},s),
\end{equation}
where
\begin{equation}
\label{6.3} {\bm W}_{\perp}({\bm k},s)= - i {\bm k} \left( {\sf I}
-\frac{{\bm k}{\bm k}}{k^{2}} \right) :{\sf R} ({\bm k},s).
\end{equation}
Substitution of the expression of ${\sf R}$ given in Eq.\
(\ref{5.22}) yields
\begin{equation}
\label{6.4} {\bm W}_{\perp}({\bm k},s)  =  -i \int_{0}^{s} d
s^{\prime} \int d{\bm c}\, {\bm k} \cdot {\bm c} \left( {\bm c} -
\frac{{\bm c} \cdot {\bm k}}{k^{2}}\, {\bm k} \right)
 \mathcal{U} ({\bm k},{\bm c},s^{\prime})
\mathcal{P}_{\perp} \widetilde{S} ({\bm k},{\bm c},s-s^{\prime}).
\end{equation}
By using this expression, it is shown in appendix \ref{ap2} that for
two components, $W_{{\perp},i} ({\bm k},s)$ and
$W_{{\perp},j} ({\bm k},s)$, of the noise of the transverse velocity field,
to lowest order in $k$ it is
\begin{equation}
\label{6.5} \langle W_{{\perp},i} ({\bm k},s)W_{{\perp},j} ({\bm
k}^{\prime},s^{\prime})\rangle_{\text{H}} = \delta_{i,j}
\delta_{{\bm k},-{\bm k}^{\prime}} \frac{\widetilde{V}^{2}}{N} k^{2}
G(|s-s^{\prime}|),
\end{equation}
for $ s>s^{\prime} \gg1$. Here $\widetilde{V}$ is the volume of the
system in the length scale $l$, i.e. $\widetilde{V} \equiv N/n
\lambda^{d}$, and
\begin{equation}
\label{6.6} G(|s|) = \int d{\bm c}\int d{\bm c}^{\prime}
\Delta_{xy}({\bm c}) \Delta_{xy} ({\bm c}^{\prime}) \widetilde{\psi}_{\text{HCS}}
({\bm c},s ; {\bm c}^{\prime}),
\end{equation}
with
\begin{equation}
\label{6.7} \widetilde{\psi}_{\text{HCS}} ({\bm c},s ; {\bm
c}^{\prime}) = \int d{\bm l}\, \widetilde{h} ({\bm l},{\bm c},s;
{\bm c}^{\prime}).
\end{equation}
The distribution $\widetilde{h} ({\bm l},{\bm c},s; {\bm
c}^{\prime})$ is defined in Eq.\ (\ref{2.26}). Then, by integration
of Eq.\ (\ref{2.27}) it follows that $ \widetilde{\psi}_{\text{HCS}}
({\bm c},s ; {\bm c}^{\prime}) $ obeys the equation
\begin{equation}
\label{6.8} \left[ \frac{\partial}{\partial s} - \Lambda ({\bm c})
\right]  \widetilde{\psi}_{\text{HCS}} ({\bm c},s ; {\bm c}^{\prime}) =0,
\end{equation}
valid for $s>0$. In principle, this equation must be solved with the
initial condition $ \widetilde{\psi}_{\text{HCS}} ({\bm c}; {\bm
c}^{\prime})$, given by
\begin{equation}
\label{6.9}  \widetilde{\psi}_{\text{HCS}} ({\bm c}; {\bm
c}^{\prime}) =  \int d{\bm l}\, \widetilde{g}({\bm l},{\bm c}, {\bm
c}^{\prime})+ \delta ({\bm c}-{\bm c}^{\prime}) \chi({c}),
\end{equation}
obtained by integration of Eq.\ (\ref{2.28}). Nevertheless, it has
been shown in ref \cite{ByR04} by particle simulation that
contributions from the correlations in the HCS of dilute granular
gases are negligible, at least for not too strong inelasticity,  $\alpha \agt 0.5$. Therefore, the
term involving $\widetilde{g}$ in Eq.\ (\ref{6.9}) has been
neglected in the results reported below.

The solution of Eq. (\ref{6.2}) in the limit of large $s$ can be
written as
\begin{equation}
\label{6.10} \delta \omega_{\perp,i} ({\bm k},s)= \int_{-\infty}^{s}
ds_{1}\, e^{\lambda_{\perp}(k)(s-s_{1})} W_{\perp,i} ({\bm k},s),
\end{equation}
where
\begin{equation}
\label{6.11} \lambda_{\perp}(k) \equiv
\frac{\zeta_{0}}{2}-\widetilde{\eta} k^{2}.
\end{equation}
Next, using Eq. (\ref{6.5}) it is obtained that
\begin{equation}
\label{6.12} \langle\delta \omega_{\perp,i}({\bm k},s) \delta
\omega_{\perp,i} ({\bm k}^{\prime},s)\rangle_{\text{H}} = -
\frac{\widetilde{V}^{2}}{2N}\, k^{2}\delta_{{\bm k},-{\bm
k}^{\prime}} \frac{\widetilde{\eta}^{\prime}}{ \lambda_{\perp}(k)},
\end{equation}
with the coefficient $\widetilde{\eta}^{\prime}$ defined as
\begin{equation}
\label{6.13} \widetilde{\eta}^{\prime} = 2 \int_{0}^{\infty} ds\,
G(|s|) e^{\frac{\zeta_{0}}{2}}.
\end{equation}
This coefficient can be computed in the first Sonine approximation.
Some details of the calculations are given in appendix \ref{ap3}.
The result reads
\begin{equation}
\label{6.14} \widetilde{\eta}^{\prime} = \frac{1+a_{2}(\alpha)}{8
\widetilde{\nu}(\alpha) - 3 \zeta_{0}(\alpha)},
\end{equation}
where $\widetilde{\nu}(\alpha)$ was defined in Eq.\ (\ref{5b.26}).

The two-time self-correlation function of the transverse velocity
field can also be computed from Eqs.\ (\ref{6.5}) and (\ref{6.10}).
Again, details of the calculations are given in appendix \ref{ap3}.
For $s-s^{\prime}
>0$, it is obtained
\begin{equation}
\label{6.16} \langle\delta \omega_{\perp,i}({\bm k},s) \delta
\omega_{\perp,j}({\bm k}^{\prime},s^{\prime}) \rangle_{H} \simeq -
\frac{\widetilde{V}^{2}}{2 N}\, \delta_{i,j} \delta_{{\bm k},-{\bm
k}^{\prime}} \frac{ \left( \widetilde{\eta}^{\prime} +
\widetilde{\eta}_{1} \right)k^{2}}{ \lambda_{\perp}(k)}\,
e^{\lambda_{\perp}(k) (s-s^{\prime})},
\end{equation}
where
\begin{equation}
\label{6.17} \widetilde{\eta}_{1} = \int_{0}^{\infty} ds\, G(|s|)
\left[ e^{- \lambda_{\perp}(k)s} - e^{\lambda_{\perp}(k) s} \right]
\end{equation}
It is worth to stress that this result only holds after a transient
interval $s-s^{\prime}$, and for this reason it does not reduce to
Eq.\ (\ref{6.12}) for $s=s^{\prime}$. On the other hand, the
coefficient $\widetilde{\eta}_{1}$ can be expected to be small, since
the second factor in the integrand remains small for the decay of
the first one.

\begin{figure}
\includegraphics[angle=0,width=.6\textwidth]{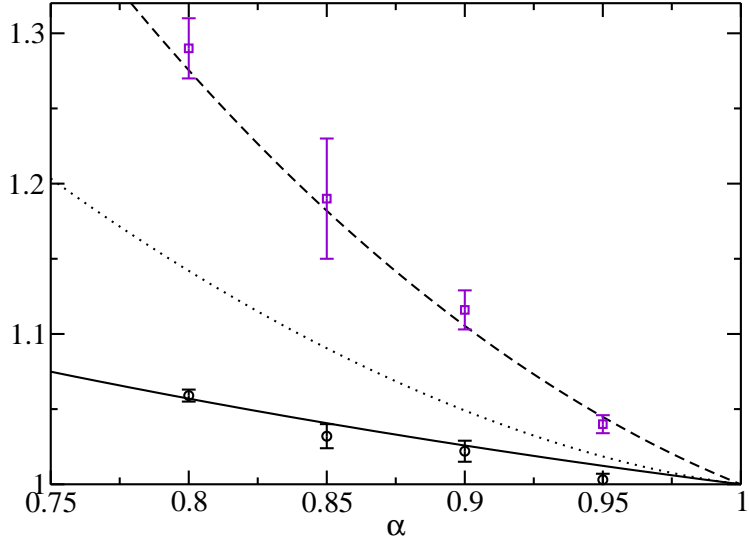}
\caption{(Color online) The shear viscosity $\widetilde{\eta}$ and
the new coefficient $\widetilde{\eta}^{\prime}$ determining the
transverse velocity fluctuations in granular gases in the HCS. The
solid and dashed lines are the theoretical predictions for
$\widetilde{\eta}$ and $\widetilde{\eta}^{\prime}$ given by Eqs.
(\protect{\ref{5a.26}}) and (\protect{\ref{6.14}}), respectively,
normalized by the elastic value of the shear viscosity,
$\widetilde{\eta}_e$. The circles
($\widetilde{\eta}/\widetilde{\eta}_e$) and squares
($\widetilde{\eta}^{\prime}/\widetilde{\eta}_e$) are molecular
dynamics simulation results. The dotted line is the result obtained
in the white noise approximation, Eq.\ (\protect{\ref{7.3}}).
\label{fig1}}
\end{figure}

To check the theory developed along the paper, Eqs. (\ref{6.16}) and
(\ref{6.12}) has been used to measure the shear viscosity
$\widetilde{\eta}$ and $\widetilde{\eta}^{\prime}$ by means of
molecular dynamics simulation of dilute granular gases. The results
turn out to be in qualitative agreement with the  theory, in the
sense that the scaled one-time correlation function is independent
of the variable $s$, and the two time correlation function only
depends on the difference $s-s^{\prime}$ and decays exponentially
after a short transient period. More details of the simulation
method employed and the analysis of the data is given in
\cite{BGyM08}. The comparison between the values of
$\widetilde{\eta}$ and $\widetilde{\eta}^{\prime}$, obtained from
the simulation results and the theoretical predictions given by
Eqs.\ (\ref{5a.26}) and (\ref{6.14}), respectively is shown in
Fig. \ref{fig1}. Of course, all the simulation data have been
obtained with low density systems in the HCS. It can be observed
that the agreement is very good over a quite range of values of the
restitution coefficient $\alpha$, then providing a very strong
support for both the theory developed here and the specific
algorithm used to compute the coefficients.

\section{Summary and discussion}
\label{s7} The primary objective here has been to investigate
hydrodynamic fluctuations in the homogeneous cooling state of dilute
granular gases, modeled as an ensemble of inelastic hard particles.
From this point of view, the fluctuating balance equations
(\ref{4.7})-(\ref{4.9}), together with the fluctuating Boltzmann
equation (\ref{3.10}) provide a solid starting point. The remaining
task is to construct explicit expressions for the fluctuating flux
and the cooling rate in terms of the (fluctuating) hydrodynamic
fields by using, for instance, the Chapman-Enskog procedure. This
part of the analysis turns out to be technically rather complex, and
has been limited here to the particular case of the transverse flow
field.

The structure of the fluctuating balance equations is similar to
those for elastic, molecular systems with two main differences, both
in the equation for the energy, Eq.\ (\ref{4.9}). The equation
contains a term, $\delta \zeta_{0}$, associated to the fluctuations
of the cooling rate  and also an intrinsic noise term
$\widetilde{S}_{\epsilon}$. Both give contributions even to zeroth
order in the gradients and therefore play a relevant role in
describing the fluctuations of global properties of the system
\cite{BGMyR04}.

With regards to the fluctuating transverse velocity field, it has
been found that it can be described by a Langevin equation, but
exhibiting two crucial differences as compared with the elastic
case. The form of the fluctuation dissipation relation changes both
qualitatively and quantitatively. The second moment of the noise
term is not determined by the shear viscosity. In addition, the
noise is not white, i.e. it presents memory effects. Both aspects
have been confirmed by the results obtained by molecular dynamics
simulations.

It could be wondered at what extension the memory effects mentioned
above are relevant. Suppose the hypothesis of a white noise would
have made and Eq. (\ref{6.5}) were substituted by
\begin{equation}
\label{7.1} \langle W_{{\perp},i} ({\bm k},s)W_{{\perp},j} ({\bm
k}^{\prime},s^{\prime})\rangle_{\text{H}} = \delta_{i,j}
\delta_{{\bm k},-{\bm k}^{\prime}} \frac{\widetilde{V}^{2}}{N} k^{2}
\widetilde{\eta}^{\prime \prime} \delta (s-s^{\prime}),
\end{equation}
with
\begin{equation}
\label{7.2} \widetilde{\eta}^{\prime \prime} = 2 \int_{0}^{\infty}
ds\, G(|s|).
\end{equation}
By using the same method as outlined in appendix \ref{ap3} it is
found that
\begin{equation}
\label{7.3} \widetilde{\eta}^{\prime \prime} = \frac{1
+a_{2}(\alpha)}{ 8 \widetilde{\nu}(\alpha)-2 \zeta_{0}(\alpha)}.
\end{equation}
This coefficient is also plotted in Fig.\ \ref{fig1}, and it is seen
to clearly underestimate the amplitude of the second moment of the
noise measured in the simulation. It is worth to stress that the
violation of the elastic fluctuation-dissipation relations is
already significant for values of the restitution coefficient
$\alpha$ of the order of $0.95$.

\section{Acknowledgements}

This research was supported by the Ministerio de Educaci\'{o}n y
Ciencia (Spain) through Grant No. FIS2008-01339 (partially financed
by FEDER funds). M.I.G.S. acknowledges financial support from Becas de
la Fundaci{\'o}n La Caixa and the French Government .

\appendix

\section{Derivation of the fluctuating hydrodynamic equations}
\label{ap1} In this appendix the derivation of Eqs.\
(\ref{4.7})-(\ref{4.9}) will be outlined. The calculations are
facilitated by using that, for arbitrary functions $f({\bm
c}_{1},{\bm c}_{2})$ and $g({\bm c}_{1},{\bm c}_{2})$, it is
\begin{eqnarray}
\label{ap1.1} \int &d{\bm c}_{1}& \int d{\bm c}_{2}\, f({\bm
c}_{1},{\bm c}_{2})\overline{T}_{0}({\bm c}_{1},{\bm c}_{2}) g({\bm
c}_{1},{\bm c}_{2}) \nonumber \\
& = & \int d{\bm c}_{1} \int d{\bm c}_{2}\, g({\bm c}_{1},{\bm
c}_{2})T_{0}({\bm c}_{1},{\bm c}_{2}) f({\bm c}_{1},{\bm c}_{2}),
\nonumber \\
\end{eqnarray}
where
\begin{equation}
\label{ap1.2} T_{0}({\bm c}_{1},{\bm c}_{2})= \int d\widehat{\bm
\sigma}\, \Theta ({\bf c}_{12} \cdot \widehat{\bm \sigma}) {\bm
c}_{12} \cdot \widehat{\bm \sigma} \left[ b_{\bm \sigma} ({\bm
c}_{1},{\bm c}_{2})-1 \right].
\end{equation}
The operator $b_{{\bm \sigma}}$ is the inverse of $b_{{\bm
\sigma}}^{-1}$ defined in Eq.\ (\ref{2.7}). From Eqs.\ (\ref{4.4}) and (\ref{3.10}),
\begin{equation}
\label{ap1.3} \frac{\partial}{\partial s} \delta \rho ({\bm l},s)
=  \int d{\bm c}\, \frac{\partial}{\partial s} \delta
\widetilde{F}_{1} ({\bm l}, {\bm c}, s) =  - \int d{\bm c}\, {\bm c} \cdot \frac{\partial}{\partial {\bm
l}}\,  \delta \widetilde{F}_{1} ({\bm l}, {\bm c}, s) + \int d{\bm
c}\, \Lambda ({\bm c})\delta \widetilde{F}_{1} ({\bm l}, {\bm c}, s)
+ \widetilde{S}_{\rho}({\bm l},s),
\end{equation}
with
\begin{equation}
\label{ap1.4} \widetilde{S}_{\rho}({\bm l},s) \equiv \int d{\bm c}\,
\widetilde{S} ({\bm l},{\bm c}, s).
\end{equation}
By using the definition given in Eq. (\ref{2.25}) and the property
(\ref{ap1.1}) it is easy to see that
\begin{equation}
\label{ap1.5} \int d{\bm c}\, \Lambda ({\bm c})\delta
\widetilde{F}_{1} ({\bm l}, {\bm c}, s) = 0.
\end{equation}
Moreover,
\begin{eqnarray}
\label{ap1.6} \langle \widetilde{S}_{\rho}({\bm l},s)
\widetilde{S}_{\rho}({\bm l}^{\prime},s^{\prime})\rangle_{\text{H}} & =
& \int d{\bm c}_{1} \int d{\bm c}_{2} \langle \widetilde{S} ({\bm
l},{\bm c}_{1}, s)
\widetilde{S} ({\bm l},{\bm c}_{2}, s)\rangle_{\text{H}} \nonumber \\
&  & = \delta (s-s^{\prime}) \delta ({\bm l}-{\bm l}^{\prime})
n^{-1} \lambda^{-d} \int d{\bm c}_{1} \int d{\bm c}_{2}\,
\widetilde{\Gamma}
({\bm c}_{1}, {\bm c}_{2}), \nonumber \\
\end{eqnarray}
where Eqs.\ (\ref{3.20}) and (\ref{3.21}) have been employed. But,
use again of Eq.\ (\ref{ap1.1}) yields
\begin{equation} \label{ap1.7}
\int d{\bm c}_{1} \int d{\bm c}_{2}\, \widetilde{\Gamma} ({\bm c}_{1}, {\bm
c}_{2})=0
\end{equation}
and consequently the second moment of $\widetilde{S}_{\rho}$
vanishes implying that the noise itself also identically vanishes.
When all the above results are substituted into Eq.\ (\ref{ap1.3})
and the definition in Eq.\ (\ref{4.5}) is taken into account, Eq.
(\ref{4.7}) follows directly.

The balance equation for the fluctuating velocity field is obtained
in a similar way. Multiplication of Eq. (\ref{3.10}) by ${\bm
c}_{1}$ and integration over it gives
\begin{equation}
\label{ap1.8} \frac{\partial}{\partial s}\, \delta {\bm \omega}({\bm
l}_{1},s)  =   - \frac{\partial}{\partial {\bm l}_{1}} \cdot \ \delta
{\sf \Pi} ({\bm l}_{1},s) + \int d{\bm c}_{1}\, {\bm c}_{1} \Lambda
({\bm c}_{1}) \delta \widetilde{F}_{1} ({\bm l}_{1},{\bm c}_{1},s)
\widetilde{\bm S}_{\omega} ({\bm l}_{1},s),
\end{equation}
where ${\sf \Pi}$ is defined in Eq.\ (\ref{4.10}) and
\begin{equation}
\label{ap1.9} \widetilde{\bm S}_{\omega}({\bm l}_{1},s) \equiv \int
d{\bm c}_{1}\, {\bm c}_{1}  \widetilde{S} ({\bm l}_{1},{\bm
c}_{1},s).
\end{equation}
A simple calculation, using Eq. (\ref{ap1.1}) and taking into
account the momentum conservation in collisions leads to
\begin{equation}
\label{ap1.10} \int d{\bm c}_{1}\, {\bm c}_{1} \Lambda ({\bm c}_{1})
\delta \widetilde{F}_{1} ({\bm l}_{1},{\bm c}_{1},s) =
\frac{\zeta_{0}}{2}\, \delta {\bm \omega} ({\bm l}_{1},s).
\end{equation}
For the correlation of the noise $\widetilde{\bm S}_{\omega}$ by
means of Eq. (\ref{3.20}) it is found
\begin{eqnarray}
\label{ap1.11}  \langle \widetilde{\bm S}_{\omega}({\bm l}_{1},s)
\widetilde{\bm S}_{\omega}({\bm l}_{1}^{\prime},s) \rangle_{\text{H}}
& = &  -\delta (s-s^{\prime}) \delta ({\bm l}_{1}-{\bm
l}^{\prime}_{1}) n^{-1} \lambda^{-d} \int d{\bm c}_{1} \int d{\bm
c}_{1}^{\prime} {\bm c}_{1} {\bm c}^{\prime}_{1}
\left\{\left[ \Lambda ({\bm c}_{1})+ \Lambda ({\bm c}_{1}^{\prime}) \right]
\right. \nonumber \\
&& \left. \times \delta ({\bm c}_{1}-{\bm c}_{1}^{\prime}) \chi
(c_{1})  \overline{T}_{0}({\bm c}_{1},{\bm c}_{1}^{\prime})
\chi(c_{1}) \chi({c}_{1}^{\prime})\right\}.
\end{eqnarray}
The calculation of the first term on the right hand side of the
above equation is straightforward giving
\begin{equation}
\label{ap1.12}
\int d{\bm c}_{1} \int d{\bm c}_{1}^{\prime} {\bm c}_{1} {\bm
c}^{\prime}_{1}  \left[ \Lambda ({\bm c}_{1})+ \Lambda ({\bm
c}_{1}^{\prime}) \right] \delta ({\bm c}_{1}-{\bm c}_{1}^{\prime})
\chi(c_{1})= \zeta_{0}\, \int d{\bm c}_{1}\, {\bm c}_{1} {\bm
c}_{1} \chi(c_{1}) = \frac{\zeta_{0}}{2}\, {\sf I},
\end{equation}
where ${\sf I}$ is the unit tensor of dimension $d$. The evaluation
of the last term on the right hand side of Eq. (\ref{ap1.11})
involves calculating several standard angular integrals. The result
is
\begin{equation}
\label{ap1.13} \int d{\bm c}_{1} \int d{\bm c}_{1}^{\prime} {\bm
c}_{1} {\bm c}^{\prime}_{1} \overline{T}_{0}({\bm c}_{1},{\bm
c}_{1}^{\prime})
\chi(c_{1}) \chi({c}_{1}^{\prime})
= \frac{(1-\alpha^{2})\pi^{\frac{d-1}{2}}}{4 \Gamma \left(
\frac{d+3}{2}\right)d}\, {\sf I} \int d{\bm c}_{1} \int d {\bm
c}_{1}^{\prime} c_{11^{\prime}}^{3} \chi(c_{1})
\chi(c_{1}^{\prime}) = \frac{\zeta_{0}}{2}\ {\sf I}.
\end{equation}

Substitution of Eqs.\, (\ref{ap1.12}) and (\ref{ap1.13}) into Eq.
(\ref{ap1.11}) shows that the second moment of the noise ${\bm
S}_{\omega}$ vanishes implying that the noise itself also vanish
identically. Then Eqs.\ (\ref{ap1.8}) and (\ref{ap1.10}) lead to
Eq.\ (\ref{4.8}).

Finally, multiplication of Eq.\ (\ref{3.10}) by $2c_{1}^{2}/d$ and
integration over ${\bm c}_{1}$ yields
\begin{eqnarray}
\label{ap1.14} \frac{\partial}{\partial s}\, \delta \epsilon ({\bm
l}_{1},s) & = &  - \frac{2}{d}\, \frac{\partial}{\partial {\bm
l}_{1}}\, \cdot \delta {\bm \phi} ({\bm l}_{1},s)+ \frac{d+2}{d}\,
\frac{\partial}{\partial {\bm l}_{1}} \cdot \delta
\omega ({\bm l}_{1},s) \nonumber \\
&& + \frac{2}{d} \int d{\bm c}_{1}\, c_{1}^{2} \Lambda ({\bm c}_{1})
\delta \widetilde{F}_{1} ({\bm l}_{1},{\bm c}_{1},s)+
\widetilde{S}_{\epsilon}({\bm l}_{1},s),
\end{eqnarray}
with the heat flux $\delta {\bm \phi}$ defined by Eq.\ (\ref{4.11}).
Given that the kinetic energy is not conserved in collisions, the
calculation of the second term on the right hand side of the above
equation is more involved than in previous cases, although it is
still quite easy to get
\begin{equation}
\label{ap1.15}\frac{2}{d} \int d{\bm c}_{1}\, c_{1}^{2} \Lambda
({\bm c}_{1}) \delta \widetilde{F}_{1} ({\bm l}_{1},{\bm c}_{1},s) =
- \delta \zeta_{0}({\bm l},s)+ \zeta_{0} \delta \epsilon ({\bm
l},s).
\end{equation}
The expression of the fluctuating cooling rate $\delta \zeta_{0}$ is
given in Eq.\ (\ref{4.14}).

Also the calculation of the second moment of the noise term
$\widetilde{S}_{\epsilon}$ is more cumbersome than for the other
ones. Actually, it is the only one that does not vanish. The result
is provided by Eq.\ (\ref{4.16}) with
\begin{eqnarray}
\label{ap1.16} \Phi({\bm c}_{1},{\bm c}_{2}) & \equiv &
\frac{\pi^{\frac{d-1}{2}}}{\Gamma \left( \frac{d+5}{2} \right)
d^{2}} \left\{ \left[ d+1-4 \alpha -(d+5)\alpha^{2} \right] C^{2}
c_{12}^{3} + \frac{(1-\alpha^{2})(d+5-2 \alpha^{2})}{4}\ c_{12}^{5}
\right. \nonumber \\
&& \left. +2(1+\alpha)(2d+3-3\alpha) c_{12} ({\bm C} \cdot {\bm c}_{12})^{2}
\right\},
\end{eqnarray}
where ${\bm c}_{12} \equiv {\bm c}_{1} -{\bm c}_{2}$ and ${\bm C}
\equiv ({\bm c}_{1}+{\bm c}_{2})/2$.

\section{Derivation of Eq.\ (\protect{\ref{6.5}})}
\label{ap2} The Fourier transform of Eq.\ (\ref{3.20}) is
\begin{equation}
\label{ap2.1} \langle\widetilde{S} ({\bm k},{\bm c},s) \widetilde{S}
({\bm k}^{\prime},{\bm c}^{\prime},s^{\prime}) \rangle_{\text{H}} =
\delta_{{\bm k}, -{\bm k}^{\prime}} \delta(s-s^{\prime})
\frac{\widetilde{V}^{2}}{N} \widetilde{\Gamma} ({\bm c},{\bm
c}^{\prime}).
\end{equation}
where $\widetilde{V} \equiv V \lambda^{-d}$, $V$ being the volume
of the system. Then, for any two components of ${\bm W}_{\perp}$,
from Eq. (\ref{6.4}) it
is obtained that
\begin{eqnarray}
\label{ap2.2}  \langle  W_{\perp,i} ({\bm k},s)W_{\perp,j} ({\bm
k}^{\prime},s^{\prime})\rangle_{\text{H}}
& = &\frac{\widetilde{V}^{2}}{N} \delta_{ij} \delta_{{\bm k}-{\bm
k}^{\prime}} k^{2} \int_{0}^{s} ds_{1} \int_{0}^{s^{\prime}}
ds_{1}^{\prime} \int d{\bm c} \int d{\bm c}^{\prime} c_{\parallel}
c_{\perp,i} c^{\prime}_{\parallel} c^{\prime}_{\perp,j}
 \mathcal{U} ({\bm k},{\bm c},s_{1}) \nonumber \\
 && \times \mathcal{U} (-{\bm
k},{\bm c}^{\prime},s_{1}^{\prime}) \delta
(s-s_{1}-s^{\prime}+s^{\prime}_{1})
\mathcal{P}_{\perp} \mathcal{P}^{\prime}_{\perp} \widetilde{\Gamma}
({\bm c},{\bm c}^{\prime}).
\end{eqnarray}
Here $\mathcal{P}^{\prime}$ is defined like $\mathcal{P}$, but
acting on functions of the velocity ${\bm c}^{\prime}$, and
$c_{\parallel}$ denotes de component of ${\bm c}$ along ${\bm k}$.
Now, it is taken into account that: i) Because calculations are
restricted to the lowest order in $k$, which is $k^{2}$, the $k$
factors in the $\mathcal{U}$ operators can be neglected, i.e. the
operator $L({\bm k},{\bm c})$ in Eq.\ (\ref{5.16}) can be replaced
by $\Lambda ({\bm c})$, and ii), Eq. (\ref{5.12}) implies that, in
particular, $\langle\chi c_{i}c_{j}|\xi_{\beta}\rangle=0$;
$\beta=1,\ldots,d+2$, for $i\neq j$. Then, the projection operator
$\mathcal{P}_{\perp}$ can be eliminated everywhere in Eq.\
(\ref{ap2.2}). Note that this can be only done after the operator
$L({\bm k},{\bm c})$ has been replaced by $\Lambda({\bm c})$. Then
by carrying out the integral over $s_{1}$, it is obtained
\begin{eqnarray}
\label{ap2.3} \langle W_{\perp,i} ({\bm k},s)W_{\perp,j} ({\bm
k}^{\prime},s^{\prime})\rangle_{\text{H}} & = &
 {\frac{\widetilde{V}^{2}}{N}}\, \delta_{{\bm k},-{\bm
k}^{\prime}} k^{2} \int d {\bm c}\int d{\bm c}^{\prime}\,
c_{\parallel}c_{\perp,i}c^{\prime}_{\parallel} c^{\prime}_{\perp,j}
e^{ (s-s^{\prime}) \Lambda ({\bm c})}  \nonumber \\
&& \times \int_{0}^{s^{\prime}} d s^{\prime}_{1}\, e^{s^{\prime}_{1}
\left[ \Lambda ({\bm c})+\Lambda ({\bm c}^{\prime}) \right]} \widetilde{\Gamma}
({\bm c},{\bm c}^{\prime}),
\end{eqnarray}
for $s>s^{\prime} $. Again, it is assumed that the HCS is linearly
stable with respect to homogeneous perturbations for the system
under consideration. Then, taking the limit $s^{\prime}\gg1 $, the
above equation can be rewritten as
\begin{equation}
\label{ap2.4} \langle W_{\perp,i} ({\bm k},s)W_{\perp,j}
({\bm k}^{\prime},s^{\prime})\rangle_{\text{H}}
= {\frac{\widetilde{V}^{2}}{N}}\, \delta_{{\bm k},-{\bm
k}^{\prime}} k^{2} \int d {\bm c}\int d{\bm c}^{\prime}\,
c_{\parallel}c_{\perp,i}c^{\prime}_{\parallel} c^{\prime}_{\perp,
j}  \times e^{ (s-s^{\prime}) \Lambda ({\bm c})} \widetilde{
\vartheta}_{\text{HCS}}({\bm c},{\bm c}^{\prime}).
\end{equation}
The function $\widetilde{\vartheta}_{\text{HCS}}({\bm c},{\bm
c}^{\prime})$ is the solution of the equation
\begin{equation}
\label{ap2.5} \left[ \Lambda({\bm c})+\Lambda ({\bm c}^{\prime})
\right] \widetilde{\vartheta}_{\text{HCS}}({\bm c},{\bm c}^{\prime}) = - \widetilde{\Gamma} ({\bm
c},{\bm c}^{\prime}),
\end{equation}
i.e., (see Eq. (\ref{2.29})),
\begin{equation}
\label{ap2.6} \widetilde{\vartheta}_{\text{HCS}}({\bm c},{\bm
c}^{\prime})= \int d{\bm l} \widetilde{h}({\bm l},{\bm c};{\bm
c}^{\prime}).
\end{equation}
To give account of the time dependence in Eq. (\ref{ap2.4})
introduce
\begin{equation}
\label{ap2.7} \widetilde{\psi}_{HCS} ({\bm c},s;{\bm c}^{\prime}) =
\int d{\bm l}\, \widetilde{h}({\bm l},{\bm c},s;{\bm c}^{\prime}).
\end{equation}
It obeys the equation (see Eq.\ (\ref{2.27}))
\begin{equation}
\label{ap2.8} \left[ \frac{\partial}{\partial s} - \Lambda ({\bm c})
\right]\widetilde{\psi}_{HCS} ({\bm c},s;{\bm c}^{\prime}) =0.
\end{equation}
By writing Eq.\ (\ref{ap2.4}) in terms of $\widetilde{\psi}_{HCS}
({\bm c},s;{\bm c}^{\prime})$, Eq.\ (\ref{6.5}) follows after a
trivial change in the notation, taking into account the isotropy of
$\Lambda ({\bm c})$.

\section{Some details of the calculations described in Sec. \ref{s6}}
\label{ap3} The explicit expression of the function $G(|s|)$ defined in
Eq.\ (\ref{6.6}) is
\begin{equation}
\label{ap3.1} G(|s|) =  \int d {\bm c} \int d{\bm c}^{\prime}
c_{x} c_{y}c^{\prime}_{x} c^{\prime}_{y} e^{s \Lambda ({\bm c})}
\chi (c) \delta
({\bm c}-{\bm c}^{\prime})
 =  \int d {\bm c} c_{x} c_{y} e^{s \Lambda ({\bm c})} \chi (c)
c_{x}c_{y}.
\end{equation}
The term involving the two-particle correlation function
$g_{\text{HCS}}$ in Eq. (\ref{6.9}) has been neglected. The reason
for this is that numerical simulations of dilute granular gases
\cite{ByR04} has shown that it gives much smaller contributions than
the term kept. Substitution of the above expression in Eq.\
(\ref{6.13}) and integration over time $s$ leads to
\begin{equation}
\label{ap3.2}
\widetilde{\eta}^{\prime} = 2 \int d{\bm c} c_{x} c_{y} C_{xy} ({\bm c}),
\end{equation}
where $ C_{xy} ({\bm c})$ obeys the equation
\begin{equation}
\label{ap3.3} \left[ \Lambda ({\bm c}) + \frac{\zeta_{0}}{2} \right]
C_{xy} ({\bm c})= - \chi ({c}) c_{x}c_{y}.
\end{equation}
This equation will be now solved in the first Sonine approximation, i.e.
by keeping only the first term of the expansion of the solution in Sonine
polynomials \cite{RyL77}, that due the isotropy of the operator $\Lambda ({\bm c})$
has the form
\begin{equation}
\label{ap3.4}
C_{xy}({\bm c}) = A c_{x} c_{y} \varphi_{0} (c),
\end{equation}
where $\varphi_{0}$ is the Gaussian
\begin{equation}
\label{ap3.5} \varphi_{0} = \pi^{-d/2} e^{-c^{2}}.
\end{equation}
To determine the constant $A$, Eq. (\ref{ap3.3}) is multiplied by $c_{x} c_{y}$ and integrated
over ${\bm c}$, after replacing $C_{xy}({\bm c})$ by its expression in Eq.\ (\ref{ap3.4}).
Moreover, the velocity distribution of the HCS is also approximated by its expression
in the first Sonine approximation, Eq.\, (\ref{2a.1}). From this point, the only remaining
task is to evaluate several integrals. Although lengthy, the calculation is straightforward and  similar
to the one needed to compute the Navier-Stokes transport coefficients \cite{BDKyS98}. The
result found for the constant $A$ reads
\begin{equation}
\label{ap3.6} A= \frac{1+a_{2}(\alpha)}{4 \widetilde{\nu}(\alpha)
-\frac{ 3\zeta_{0}}{2}},
\end{equation}
with $\widetilde{\nu} (\alpha)$ given by Eq.\ (\ref{5b.26}). Finally,
Eq. (\ref{ap3.4}) is substituted in Eq. (\ref{ap3.2}) and the result (\ref{6.14}) follows quite easily.

In order to derive an expression for the two-time correlation function of the
transverse velocity, Eqs. (\ref{6.5}) and (\ref{6.10}) are
employed. Taking $s>s^{\prime} \gg 1$,
\begin{eqnarray}
\label{ap3.7}  \langle  \delta \omega_{\perp,i}({\bm k},s) \delta
\omega_{\perp,j}({\bm k}^{\prime},s^{\prime})\rangle_{\text{H}}
& = & \frac{\widetilde{V}^{2}}{N}\, k^{2} \delta_{i,j} \delta_{{\bm
k},-{\bm k}^{\prime}}  \left[ \int_{-\infty}^{s^{\prime}} ds_{1}
\int_{-\infty}^{s^{\prime}} ds^{\prime}_{1}\, e^{\lambda_{\perp}(k)
(2s^{\prime}-s_{1}-s^{\prime}_{1})} G(|s_{1}-s^{\prime}_{1}|)
\right.
\nonumber \\
&& \left. + \int_{s^{\prime}}^{s} ds_{1} \int_{-\infty}^{s^{\prime}}
ds^{\prime}_{1}\, e^{\lambda_{\perp}(k)
(s+s^{\prime}-s_{1}-s^{\prime}_{1})} G(|s_{1}-s^{\prime}_{1}|)
\right]. \nonumber \\
\end{eqnarray}
The first integral on the right hand side is related with the
coefficient $\widetilde{\eta}^{\prime}$ computed before,
\begin{equation}
\label{ap3.8}
\int_{-\infty}^{s^{\prime}} ds_{1} \int_{-\infty}^{s^{\prime}}
ds^{\prime}_{1}\, e^{\lambda_{\perp}(k)
(2s^{\prime}-s_{1}-s^{\prime}_{1})} G(|s_{1}-s^{\prime}_{1}|)
= - \frac{\widetilde{\eta}^{\prime}}{2
\lambda_{\perp}(k)}.
\end{equation}
The other integral in Eq. (\ref{ap3.7}) can be evaluated for
instance by introducing the variables $g \equiv
s_{1}-s_{1}^{\prime}$ and $S \equiv (s_{1} +s_{1}^{\prime})/2$.
Then, the integral is decomposed into three. For large enough
$s-s^{\prime}$, the only remaining one is
\begin{equation}
\label{ap3.9}
\int_{0}^{\infty} dg\ G(|g|)
\int_{s^{\prime}-\frac{g}{2}}^{s^{\prime}+\frac{g}{2}} d S\,
e^{\lambda_{\perp}(k) (s+s^{\prime}-2S)}= - \frac{\widetilde{\eta}_{1}}{2
\lambda_{\perp}(k)}\, e^{\lambda_{\perp}(k)(s-s^{\prime})},
\end{equation}
with $\widetilde{\eta}_{1}$ given by Eq.\ (\ref{6.17}). Use of Eqs.
(\ref{ap3.8}) and (\ref{ap3.9}) into Eq.\ (\ref{ap3.7}) gives Eq.\
(\ref{6.16}).

\end{document}